\documentclass[12pt]{article}

\baselineskip=\normalbaselineskip

\usepackage{mathrsfs,amsbsy,amssymb,latexsym,amsfonts,amsmath,amsthm}
\usepackage{graphicx,color}
\usepackage{simplewick}

\makeatletter
\catcode`\@=11
\@addtoreset{equation}{section}

\def\@seccntformat#1{\csname the#1\endcsname.~~}
\makeatother
\newcommand{\nn}{\nonumber}

\begin{document}

\begin{titlepage}
\renewcommand{\thefootnote}{\fnsymbol{footnote}}
\begin{flushright}
KUNS-2615
\end{flushright}
\vspace*{1.0cm}

\begin{center}
{\Large \bf 
Triangle-hinge models for unoriented membranes
}
\vspace{1.0cm}

\centerline{
{Masafumi Fukuma,}%
\footnote{E-mail address: 
fukuma@gauge.scphys.kyoto-u.ac.jp} 
{Sotaro Sugishita}%
\footnote{E-mail address: 
sotaro@gauge.scphys.kyoto-u.ac.jp} and
{Naoya Umeda}%
\footnote{E-mail address: 
n\_umeda@gauge.scphys.kyoto-u.ac.jp}%
}

\vskip 0.8cm
{\it Department of Physics, Kyoto University, Kyoto 606-8502, Japan}
\vskip 1.2cm 

\end{center}

%%%%%%%%%%%%%%%%%%%%%%%%%%%%%%%%%%%%%%%
\begin{abstract}
%%%%%%%%%%%%%%%%%%%%%%%%%%%%%%%%%%%%%%%

Triangle-hinge models [arXiv:1503.08812] 
are introduced to describe worldvolume dynamics of membranes. 
The Feynman diagrams consist of triangles glued together along hinges 
and can be restricted to tetrahedral decompositions 
in a large $N$ limit. 
In this paper, 
after clarifying that all the tetrahedra resulting in the original models 
are orientable, 
we define a version of triangle-hinge models 
that can describe the dynamics of {\em unoriented} membranes. 
By regarding each triangle as representing a propagation 
of an open membrane of disk topology, 
we introduce a local worldvolume parity transformation 
which inverts the orientation of triangle, 
and define unoriented triangle-hinge models 
by gauging the transformation. 
Unlike two-dimensional cases, 
this local transformation generally relates a manifold to a nonmanifold, 
but still is a well-defined manipulation 
among tetrahedral decompositions. 
We further show that matter fields can be introduced 
in the same way as in the original oriented models. 
In particular, the models will describe unoriented membranes 
in a target spacetime 
by taking matter fields to be the target space coordinates.

%%%%%%%%%%%%%%%%%%%%%%%%%%%%%%%%%%%%%%%
\end{abstract}
%%%%%%%%%%%%%%%%%%%%%%%%%%%%%%%%%%%%%%%
\end{titlepage}

\pagestyle{empty}
\pagestyle{plain}

\tableofcontents
\setcounter{footnote}{0}

%%%%%%%%%%%%%%%%%%%%%%%%%%%%%%%%%%%%%%%
%%%%%%%%%%%%%%%%%%%%%%%%%%%%%%%%%%%%%%%
\section{Introduction}
%%%%%%%%%%%%%%%%%%%%%%%%%%%%%%%%%%%%%%%
%%%%%%%%%%%%%%%%%%%%%%%%%%%%%%%%%%%%%%%

The worldvolume theory  of membranes in a spacetime 
is equivalent to a system of three-dimensional quantum gravity 
coupled to matter fields corresponding to the target space coordinates. 
One approach to treating such class of systems is 
the use of models that generate three-dimensional random volumes.     
Triangle-hinge models \cite{Fukuma:2015xja} 
are proposed as such models. 
The dynamical variables are given by a pair of $N\times N$ symmetric matrices, 
$A$ and $B$,  
and the Feynman diagrams consist of triangles glued together along their edges. 
We can restrict diagrams 
such that they represent only 
three-dimensional tetrahedral decompositions 
by taking a large $N$ limit. 
The simplest model thus obtained corresponds 
to discretized three-dimensional pure quantum gravity 
with a bare cosmological constant. 
We can further introduce extra degrees of freedom 
representing the target space coordinates. 
A prescription to introduce such matter degrees of freedom 
to triangle-hinge models is given in \cite{Fukuma:2015haa}. 
The prescription also enables us to describe various spin systems 
such as the $q$-state Potts models coupled to quantum gravity,  
and to realize colored tensor models \cite{Gurau:2009tw, Gurau:2011xp} 
in terms of triangle-hinge models.

As pointed out in \cite{Fukuma:2015xja}, 
the original triangle-hinge models generate 
only (and all of the) orientable tetrahedral decompositions. 
In this paper, we generalize the models 
such that unoriented membranes can be treated. 
We will call the obtained models {\em unoriented triangle-hinge models}. 
In the context of string theory, 
to consider unoriented models is not just interesting 
as a mathematical generalization, 
but has a physically important meaning. 
Actually, 
an unoriented superstring theory, 
type I  superstring, 
is one of (perturbatively) consistent superstring theories. 
We expect that unoriented membrane theory is also physically important.%
\footnote{%=====
 We should comment that the low energy effective theory 
 of unoriented supermembranes 
 is not the eleven-dimensional supergravity, 
 because the 3-form fields in the supergravity multiplet cannot couple 
 to unoriented membranes. 
 Nevertheless, we expect that 
 unoriented membrane theory serves as a toy model 
 to obtain a better understanding of dynamics of membranes. 
} %=============

In two-dimensional cases,  
an unoriented theory is obtained 
by gauging the worldsheet parity of an oriented theory. 
If we discretize worldsheets by triangular decompositions, 
the gauging procedure is to treat equally two ways 
to identify an edge of a triangle with that of another triangle; 
one way preserves the local orientations of two triangles 
and the other does not. 
We will define unoriented membrane theories 
by generalizing the prescription 
to three-dimensional tetrahedral decompositions. 
Roughly speaking, the unoriented models equally treat two possible ways 
to identify a triangle of a tetrahedron with that of another tetrahedron; 
one way preserves the orientation 
and the other does not.

Here, we make comments on the treatment of orientability 
in other three-dimensional random volume theories, tensor models. 
Tensor models \cite{Ambjorn:1990ge, Sasakura:1990fs, Gross:1991hx} 
are natural generalizations of matrix models to higher dimensions. 
One can introduce various kinds of tensor models \cite{Sasakura:1990fs} 
depending on how indices of rank-3 tensors are assigned to triangles 
in tetrahedral decompositions.% 
\footnote{ %=====
 There is another type of tensor model
 (called the canonical tensor model)
 which realizes the constraints
 in the canonical quantization of gravity
 \cite{Sasakura:2011sq, Sasakura:2014gia, Sasakura:2015pxa}.
 An interesting connection to random tensor networks 
 is studied in \cite{Sasakura:2014zwa, Sasakura:2015xxa}.
} %============== 
A class of models, where each index is assigned to a vertex of a triangle,  
generate only orientable tetrahedral decompositions \cite{Sasakura:1990fs}.  
Although, by modifying the models, 
we may be able to construct unoriented models, 
it is difficult to solve them.  
Analytical treatment of tensor models is improved in colored tensor models 
\cite{Gurau:2009tw, Gurau:2011xp}. 
The models only generate tetrahedral decompositions belonging 
to a specific class.%
\footnote{%=====
 See, e.g., \cite{Tanasa:2015uhr}  
 for an attempt to relax the restriction on the diagrams 
 generated in colored tensor models. 
} %=============
This restriction enables us to take a $1/N$ expansion of the free energy 
\cite{Gurau:2010ba, Gurau:2011aq}. 
Furthermore, in the so-called invariant models \cite{Bonzom:2012hw} 
one can take the double scaling limit \cite{Dartois:2013sra, Bonzom:2014oua}. 
It can be shown that tetrahedral decompositions 
generated by colored tensor models are orientable. 
However, if we try to modify the models to unoriented ones, 
the solvability of the models may be lost.

Original triangle-hinge models are expected to be solvable 
because the dynamical variables are matrices. 
In fact, for simple models 
(such as the models characterized by matrix rings), 
the interaction terms in the action 
can be rewritten to the traces of powers of matrices \cite{FSU_prep}. 
Thus, in order to reduce the systems to those of eigenvalues, 
we only need to integrate the exponential of the quadratic term in the action
over the angular parts of matrices. 
Moreover, numerical integrations \cite{FSU_prep} show that 
the eigenvalue distributions of matrices $A$ and $B$ 
have a similar structure to those of one-matrix models with double-well potentials, 
and that the effective theory of eigenvalues 
for either of matrix $A$ or $B$ has critical points. 
Since there are integration contours 
for which the matrix integrations are finite \cite{FSU_prep}, 
it is highly expected that the original oriented triangle-hinge models 
have well-defined continuum limits. 
We will see that the actions of unoriented triangle-hinge models 
have a similar structure to the original ones, 
and thus we expect that unoriented triangle-hinge models are also solvable 
and might be easier to solve due to the higher symmetry they have.

This paper is organized  as follows. 
In section \ref{sec_review}, 
we review triangle-hinge models 
and show that the tetrahedral decompositions generated by the models 
are orientable. 
In section \ref{sec_unorient}, 
after reviewing matrix models for unoriented strings, 
we define unoriented membrane theories in terms of tetrahedral decompositions. 
In section \ref{sec_unorient_th}, 
we give a version of triangle-hinge models 
that realize unoriented membrane theories. 
In section \ref{sec_matter}, 
we show that matter fields can be introduced 
also to unoriented triangle-hinge models 
by the same procedure as \cite{Fukuma:2015haa}.  
Section \ref{sec_conclusion} is devoted to conclusion.

%%%%%%%%%%%%%%%%%%%%%%%%%%%%%%%%%%%%%%%
%%%%%%%%%%%%%%%%%%%%%%%%%%%%%%%%%%%%%%%
\section{Orientability in triangle-hinge models}
\label{sec_review}
%%%%%%%%%%%%%%%%%%%%%%%%%%%%%%%%%%%%%%%
%%%%%%%%%%%%%%%%%%%%%%%%%%%%%%%%%%%%%%%

In this section, we clarify the fact 
that the original triangle-hinge model \cite{Fukuma:2015xja}
generates the set of {\em oriented} tetrahedral decompositions. 

%%%%%%%%%%%%%%%%%%%%%%%%%%%%%%%%%%%%%%%
\subsection{Brief review of triangle-hinge models}
\label{subsec_review_TH}
%%%%%%%%%%%%%%%%%%%%%%%%%%%%%%%%%%%%%%%

\begin{figure}[htbp]
\begin{quote}
\begin{center}
\includegraphics[height = 3.0cm]{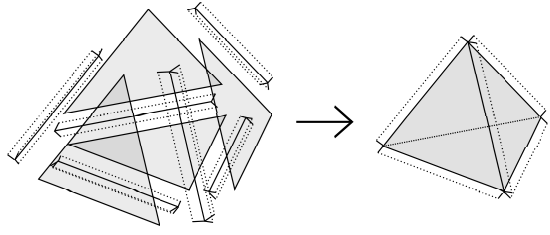}
\caption{
A part of a configuration consisting of triangles glued together 
with multiple hinges \cite{Fukuma:2015xja}.  
}
\label{tetra_triangle-hinge}
\vspace{-3ex}
\end{center}
\end{quote}
\end{figure}

Triangle-hinge models \cite{Fukuma:2015xja} are designed 
to generate Feynman diagrams 
each of which can be regarded as a collection of triangles 
glued together along multiple hinges 
and will eventually give a three-dimensional tetrahedral decomposition 
in a large $N$ limit (see Fig.~\ref{tetra_triangle-hinge}).%
\footnote{%-----
 Here, a (multiple) hinge is an object connecting edges of triangles. 
 A hinge with $k$ edges is called a $k$-hinge.
} %------------- 
%%%%%%%%%%%%%%%%%%%%%%%
%%%%%%%%%%%%%%%%%%%%%%%. 

\begin{figure}[htbp]
\begin{quote}
\begin{center}
 \includegraphics[height = 3.0cm]{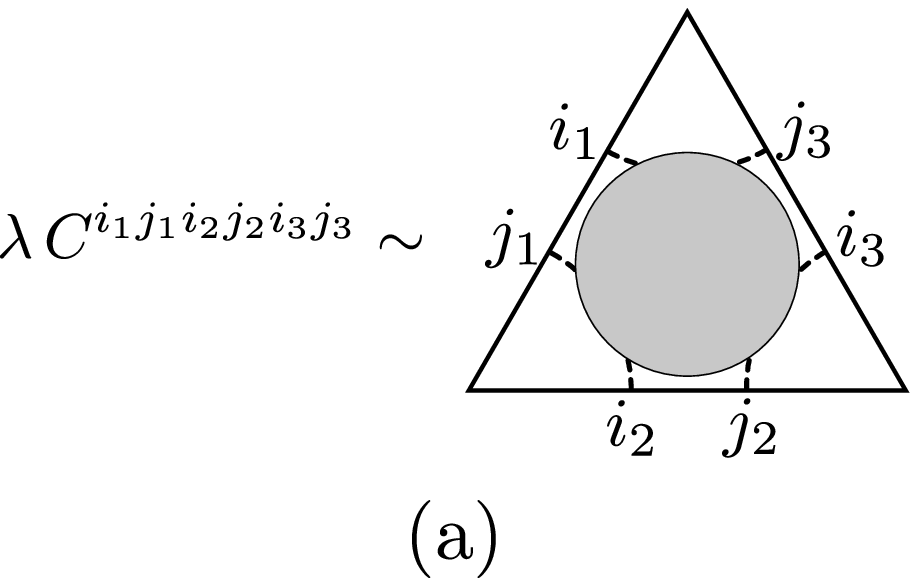}
 \hspace{0.5cm}
 \includegraphics[height = 3.0cm]{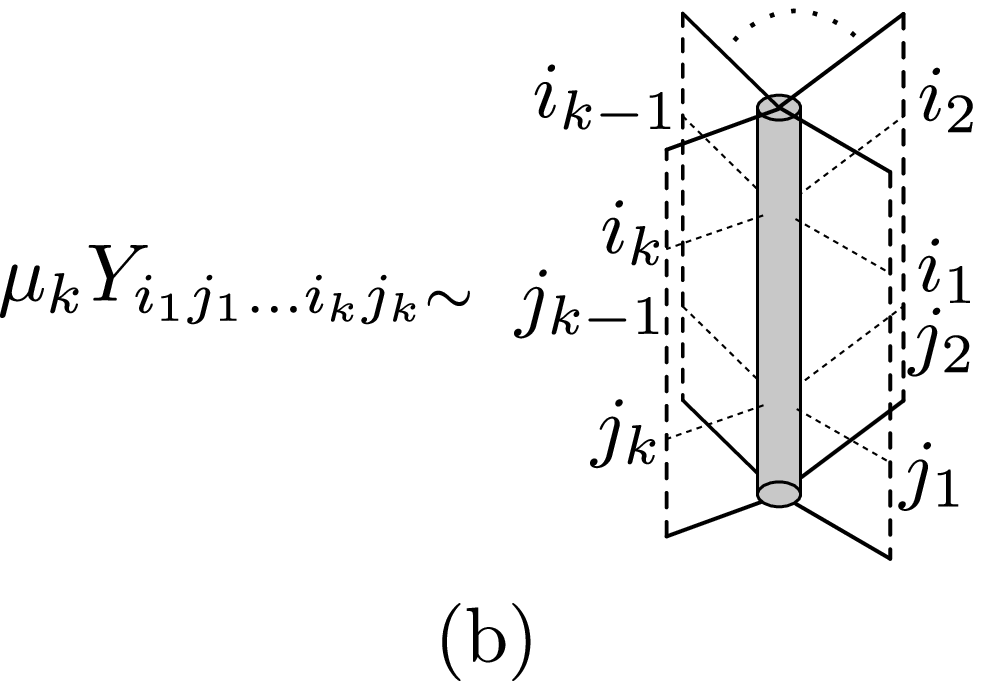}
 \caption{
 Interaction vertices corresponding to (a) triangles and (b) $k$-hinges 
 \cite {Fukuma:2015xja}. 
 }
\label{fig:triangle-hinge}
\vspace{-3ex}
\end{center}
\end{quote}
\end{figure}

\begin{figure}[htbp]
\begin{quote}
\begin{center}
 \includegraphics[height = 3.5cm]{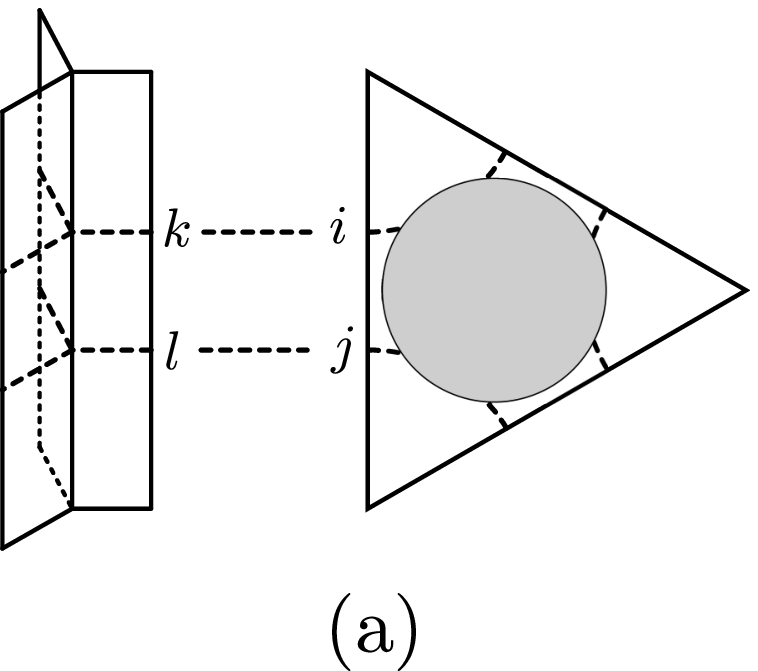}
 \hspace{1.5cm}
 \includegraphics[height = 3.5cm]{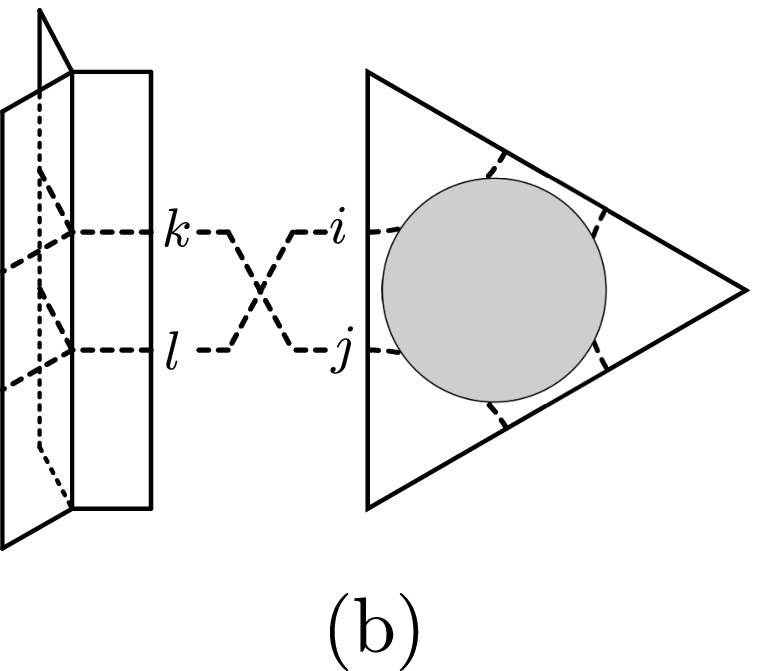}
 \caption{
 Two ways to connect an edge of a triangle to an edge of a hinge. 
 If we fix the position of the triangle, 
 the right figure (b) represents the diagram 
 where the edge of the hinge is glued to that of the triangle 
 upside down compared to the left figure (a). 
 }
\label{glue_triangle_hinge}
\vspace{-3ex}
\end{center}
\end{quote}
\end{figure}

The dynamical variables are given 
by a pair of $N\times N$ real symmetric matrices, 
$A=(A_{ij}=A_{ji})$ and $B=(B^{ij}=B^{ji})$, 
and the action takes the form%
\footnote{%-----
 Note that we have included in the action 
 the interaction term corresponding to 1-hinges.
} %-------------
\begin{align}
 S[A, B] &=  \frac{1}{2} \,[A B] -\frac{\lambda}{6} \,[CAAA]
 - \sum_{k \geq 1} \frac{\mu_k}{2k}\, 
 [Y_k \underbrace{B \cdots B}_k] 
\nn\\
 &\equiv \frac{1}{2} \,A_{ij}B^{ij}
 - \frac{\lambda}{6}\, C^{i_1j_1i_2j_2i_3j_3} 
 A_{i_1j_1}A_{i_2j_2}A_{i_3j_3} 
 - \sum_{k \geq 1} \frac{\mu_k}{2k}\, Y_{i_1 j_1 \ldots i_k j_k}
 B^{i_1 j_1} \cdots B^{i_k j_k}. 
\label{original_action_generic}
\end{align} 
The free energy is given by 
\begin{align}
 F = \log \int \!  d A \,dB \, e^{-S[A,B]}\,,
\end{align}
and if we expand $F$ with respect to $\lambda$ and  $\mu_k$, 
each term is expressed by a group of Wick contractions as usual. 
There are two types of interaction vertices; 
one (coming from $\lambda\, [C A^3]$) corresponds to a triangle 
and the other (coming from $\mu_k [Y_k B^k]$) to a multiple hinge 
(see Fig.~\ref{fig:triangle-hinge}). 
%%%%%%%%%%%%%%%%%%%%%%%
%%%%%%%%%%%%%%%%%%%%%%% 
The coefficients $C^{ijklmn}$ and $Y_{i_1 j_1\ldots i_k j_k}$ 
are real constant tensors, 
and we will not assume any symmetry 
for the indices of $C^{i_1 j_1 i_2 j_2 i_3 j_3}$ 
and $Y_{i_1 j_1 \ldots i_k j_k}$ 
until we give their explicit forms later 
[see \eqref{cplus} and \eqref{model_Y}].%
\footnote{%=====
 In fact, 
 when multiplied by $A_{i_1j_1}A_{i_2j_2}A_{i_3j_3}$ $(A_{ij}=A_{ji})$, 
 only fully symmetric part of $C^{i_1j_1i_2j_2i_3j_3}$ survive 
 that are invariant under interchanges of indices 
 $i_\alpha$ and $j_\alpha$ $(\alpha=1,\ldots,3)$ 
 and under permutations of three pairs of indices 
 $(i_1 j_1)$, $(i_2 j_2)$, $(i_3 j_3)$, 
 so we could have assumed 
 that the tensor $C$ in \eqref{original_action_generic} 
 have the symmetry  
 $C^{i_1 j_1 i_2 j_2 i_3 j_3}=C^{i_2 j_2 i_3 j_3 i_1 j_1}
 =C^{j_1 i_1 i_2 j_2 i_3 j_3}=C^{i_2 j_2 i_1 j_1 i_3 j_3}$. 
 We, however, do not assume this symmetry 
 and regard contractions  
 using $C^{i_1j_1i_2j_2i_3j_3}$, $C^{i_2 j_2 i_3 j_3 i_1 j_1}$, 
 $C^{j_1 i_1 i_2 j_2 i_3 j_3}$ or $C^{i_2j_2i_1j_1i_3j_3}$ 
 as giving independent Wick contractions \cite{Fukuma:2015xja, Fukuma:2015haa}. 
 Note that only the fully symmetric part is actually left 
 when all the diagrams are summed.  
 The same argument is applied to the hinge parts.
} %===============
They are connected by a free propagator 
% (i.e.\ a Wick contraction) 
\begin{align}
 \contraction{}{A}{_{ij}}{B} A_{ij} B^{kl}
 = \delta_i^k \delta_j^l + \delta_i^l \delta_j^k .  
\label{wick}
\end{align}
The two terms on the right-hand side in \eqref{wick} 
express that there are two ways to connect 
an edge of a hinge to an edge of a triangle 
as shown in Fig.~\ref{glue_triangle_hinge}. 
%%%%%%%%%%%%%%%%%%%%%%%
%%%%%%%%%%%%%%%%%%%%%%%
We regard the two types of pairing 
as representing two independent Wick contractions 
and write the first type as 
$\contraction{}{A}{}{B} A B^{(+)}$ or simply as $A B^{(+)}$ 
and the second type as $\contraction{}{A}{}{B} A B^{(-)}=A B^{(-)}$, 
where we have omitted the indices $i,j,\ldots$.%
\footnote{%=====
 Of course, the two pairs appear in a combined way 
 as $\contraction{}{A}{}{B}A B^{(+)} + \contraction{}{A}{}{B}A B^{(-)}$. 
} %=============
Then, a group of Wick contractions (denoted by $x$) 
can be specified uniquely in a form such as  
\begin{align}
 x = [C A_{1} A_{2} A_{3}]\, [C A_{4} A_{5} A_{6}]
 \cdots [Y_{k} B_{1}^{(+)} B_{6}^{(-)} \cdots ]\,
 [Y_{k'}  B_{5}^{(+)} \cdots] \cdots \,,
\label{a_contraction}
\end{align}
where the subscript $I$ of $A_I$ and $B_I^{(\pm)}$ 
indicates that they belong to the $I$-th contraction 
of type $\contraction{}{A}{}{B} A B^{(\pm)}$.  
As explained above, 
we do not impose a symmetry for the coefficients $C$ and $Y$, 
and think that changing the order of the labels in $[CA^3]$ or $[Y_k B^k]$ 
(e.g., replacing $[C A_1 A_2 A_3]$ by $[C A_2 A_3 A_1]$) 
leads to a different group of Wick contractions.

In \cite{Fukuma:2015xja}, we investigated in detail the case 
where the interaction terms $\mu_k [Y_k B^k]$ 
are characterized by a semisimple associative algebra $\mathcal{A}$. 
Let a basis of $\mathcal{A}$ be $\{e_i\}$ $(i=1,\ldots, N)$, 
where $N$ is the dimension of $\mathcal{A}$ as a linear space. 
The multiplication $\times$ of $\mathcal{A}$ is then specified 
by the structure constants $y_{ij}^{\phantom{ij}k}$ as
\begin{align}
 e_i \times e_j =y_{ij}^{\phantom{ij}k} e_k \,. 
\end{align}
We further introduce a rank $k$ tensor 
from the structure constants as%
\footnote{%=====
 The rank one tensor $y_{i}$ is especially defined 
 as $y_i = y_{ij}^{\phantom{ij}j}$. 
} %=============
\begin{align}
 y_{i_1 \ldots i_k} &= 
 y_{i_1 j_1}^{\phantom{i_1 i_1}j_k} y_{i_2 j_2}^{\phantom{i_2 j_2}j_1} 
 \cdots y_{i_k j_k}^{\phantom{i_k j_k}j_{k-1}},
\label{k_tensor}
\end{align}
which enjoys the cyclic symmetry, 
$y_{i_1\ldots i_k}=y_{i_2\ldots i_k i_1}$. 
Then the coupling constants $Y_{i_1 j_1\ldots i_k j_k}$ 
associated with hinges 
are defined to be
\begin{align}
 Y_{i_1 j_1\ldots i_k j_k} \equiv y_{i_1\ldots i_k} y_{j_k\ldots j_1} \,,
\label{k_tensor_algebra}
\end{align}
which enjoy the symmetry properties 
\begin{align}
 Y_{i_1 j_1 \ldots i_k j_k} = Y_{i_2 j_2 \ldots i_k j_k i_1 j_1}
 = Y_{j_k i_k \ldots j_1 i_1} \,.
\label{symmetry_Y}
\end{align}

\begin{figure}[htbp]
\begin{quote}
\begin{center}
 \includegraphics[height = 3.0cm]{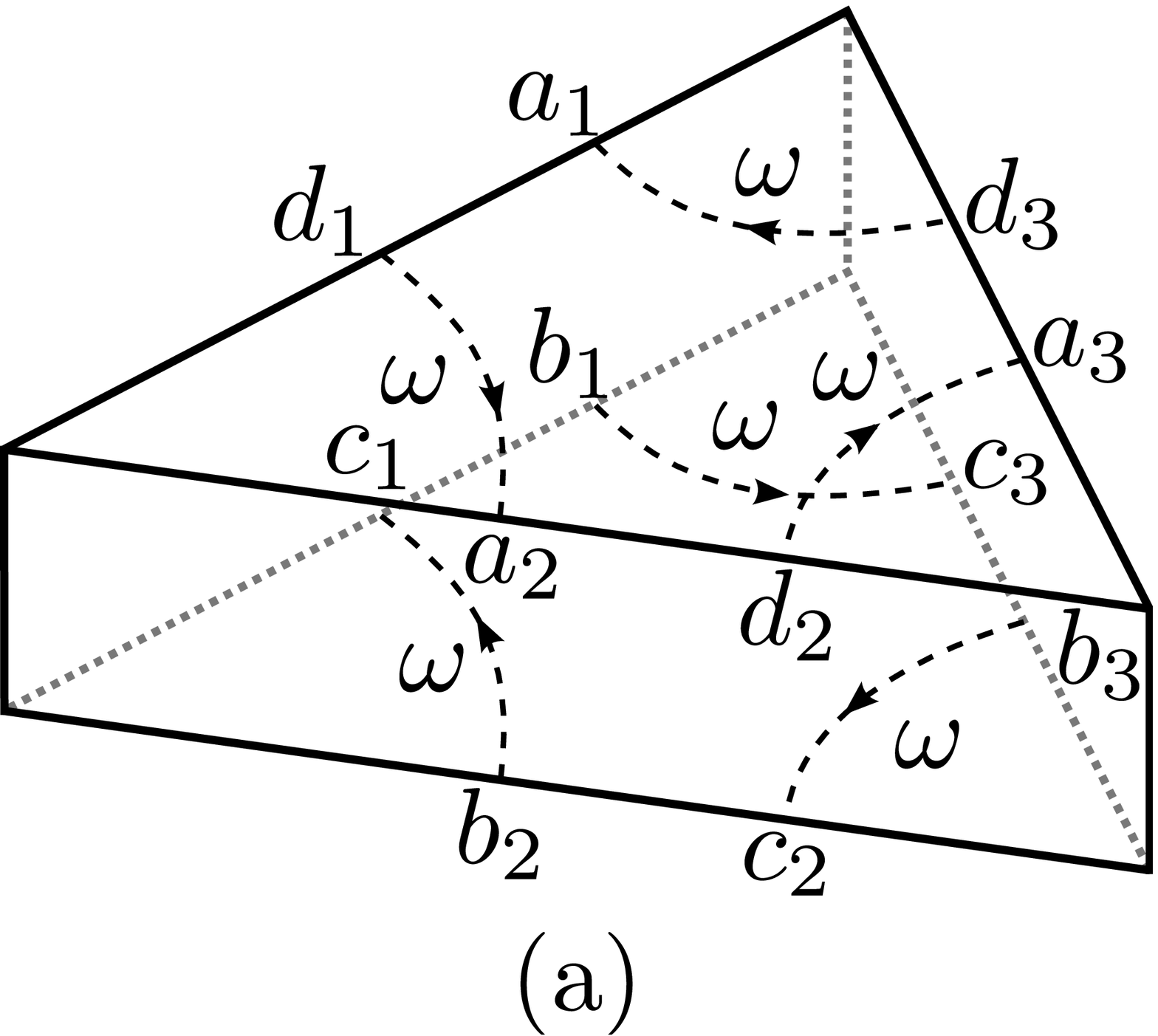}
 \hspace{2.5cm}
 \includegraphics[height = 4.0cm]{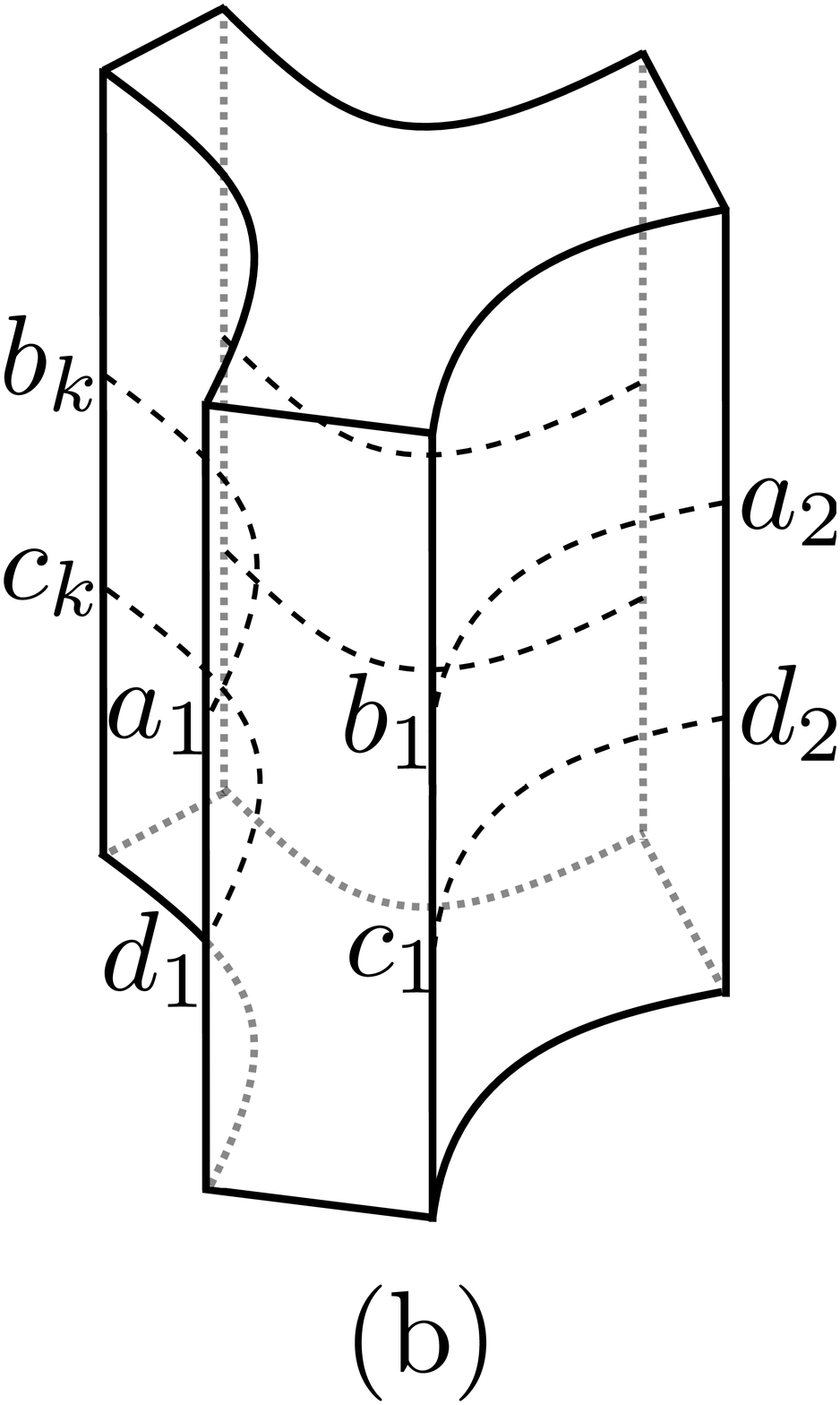}
 \caption{
 (a): thickened triangles and (b): thickened hinges \cite{Fukuma:2015xja}. 
 Each thickened triangle has six index lines 
 with arrows corresponding to $\omega$.} 
\label{thickened_vertices}
\vspace{-3ex}
\end{center}
\end{quote}
\end{figure}

\begin{figure}[htbp]
\begin{quote}
\begin{center}
 \includegraphics[height = 5.0cm]{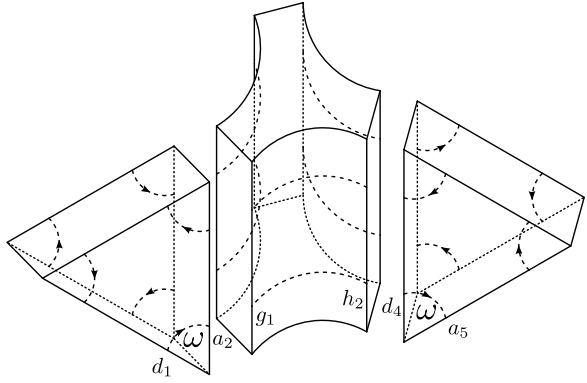}
 \caption{
 Gluing of thickened triangles along a hinge. 
 Indices $a_2$ and $d_4$ are contracted with $g_1$ and $h_2$, 
 respectively, giving $(\omega^2)^{d_1 a_5}$. 
 } 
\label{thickened_triangle-hinge_index}
\vspace{-3ex}
\end{center}
\end{quote}
\end{figure}

In order to restrict configurations 
so as to represent only tetrahedral decompositions, 
we consider the case where the algebra $\mathcal{A}$ 
is a matrix ring $\mathcal{A}=M_{n=3m}(\mathbb{R})$ 
with $n$ being a multiple of three \cite{Fukuma:2015xja}.  
The dimension of $\mathcal{A}$ is then given by $N = n^2 =(3m)^2$. 
We take a basis $\{e_i\}$ to be $\{e_{ab}\}$ $(a,b =1,\ldots n)$,  
where $e_{ab}$ are the matrix units 
whose $(c,d)$ elements are  
$(e_{ab})_{cd} =\delta_{ac} \delta_{bd}$.  
Note that indices $i$ are replaced by double indices $ab$. 
Then, the rank $k$ tensors \eqref{k_tensor} are given by 
\begin{align}
 y_{i_1 i_2 \ldots i_k} = y_{a_1 b_1,\, a_2 b_2, \ldots,\, a_k b_k}
 =  n\, \delta_{b_1 a_2} \cdots \delta_{b_{k-1} a_k}  \delta_{b_k a_1} \,, 
\end{align}
which in turn give the $k$-hinge tensor 
$Y_{i_1 j_1 i_2 j_2 \ldots i_k j_k}$ as 
\begin{align}
 Y_{a_1 b_1 c_1 d_1,\, a_2 b_2 c_2 d_2,\, \ldots,\, a_k b_k c_k d_k}
 = n^2\, \delta_{b_1 a_2} \cdots \delta_{b_{k-1} a_k}  \delta_{b_k a_1}\,
 \delta_{c_1 d_2} \cdots \delta_{c_{k-1} d_k}  \delta_{c_k d_1}\,.
\label{model_Y}
\end{align}
We further introduce a permutation matrix $\omega$ of the following form:
\begin{align}
 \omega = 
\begin{pmatrix} 
0&1_{n/3}&0 \\ 
0&0&1_{n/3} \\ 
1_{n/3}&0&0 
\end{pmatrix},
& \qquad 1_m : m\times m \, \text{unit matrix}, 
\end{align}
and set the tensor $C^{i_1 j_1 i_2 j_2 i_3 j_3}$ 
in \eqref{original_action_generic} to be 
\begin{align}
 C_{+}^{a_1 b_1 c_1 d_1,\, a_2 b_2 c_2 d_2,\, a_3 b_3 c_3 d_3} 
 \equiv \frac{1}{n^3}\, 
\omega^{d_1 a_2} \omega^{d_2 a_3} \omega^{d_3 a_1} 
 \omega^{b_3 c_2} \omega^{b_2 c_1} \omega^{b_1 c_3} . 
\label{cplus}
\end{align} 
Note that $C_+$ enjoys the symmetry properties 
\begin{align}
 C_+^{i_1 j_1 i_2 j_2 i_3 j_3} = C_+^{i_2 j_2 i_3 j_3 i_1 j_1}
 = C_+^{j_3 i_3 j_2 i_2 j_1 i_1}\,.
\label{symmetry_C}
\end{align}
The action then takes the form
\begin{align}
 S &=\frac{1}{2}\,[AB]
 - \frac{\lambda}{6}\, [C_{+}AAA] 
 - \sum_{k\geq 1} \frac{\mu_k}{2k} \,[Y_k \underbrace{B \cdots B}_k] 
\nn\\
 &\equiv  \frac{1}{2}\,A_{abcd}B_{abcd}  \nn \\ 
 &~~~ -\frac{\lambda}{6 n ^3}\omega^{d_1 a_2}\omega^{d_2 a_3}
 \omega^{d_3 a_1} \omega^{b_3 c_2}\omega^{b_2 c_1}\omega^{b_1 c_3}  
 A_{a_1 b_1 c_1 d_1} A_{a_2 b_2 c_2 d_2} A_{a_3 b_3 c_3 d_3}  \nn \\ 
 &~~~ - \sum_{k\geq 1} \frac{n^2 \mu_k}{2k} B_{a_1 a_2 b_2 b_1} \cdots
 B_{a_{k-1} a_k b_k b_{k-1}} B_{a_k a_1 b_1 b_k} \,. 
\label{oriented_THaction} 
\end{align}
The interaction vertices  can be represented 
by thickened triangles and hinges 
(see Fig.~\ref{thickened_vertices}),  
and are connected with the use of two types of Wick contractions 
between $A_{abcd}$ and $B_{efgh}$,
\begin{align}
 \contraction{}{A}{}{B} A B^{(+)}
 = \delta_{ae}\delta_{bf}\delta_{cg}\delta_{dh} \,, 
 \qquad
 \contraction{}{A}{}{B} A B^{(-)}
 = \delta_{ag}\delta_{bh}\delta_{ce}\delta_{df} \,.
\end{align}
%%%%%%%%%%%%%%%%%%%%%%%
%%%%%%%%%%%%%%%%%%%%%%%
Note that arrows are assigned to index lines on triangles 
as in Fig.~\ref{thickened_vertices}, 
and that their directions are preserved 
when two triangles are glued together along an intermediate hinge 
(see Fig.~\ref{thickened_triangle-hinge_index}). 
%%%%%%%%%%%%%%%%%%%%%%%
%%%%%%%%%%%%%%%%%%%%%%%

We here introduce some terminology 
to discriminate between a group of Wick contractions and a Feynman diagram. 
We have already clarified our rule about Wick contractions. 
We now introduce an equivalence relation 
to the set $\{x\}$ of groups of Wick contractions, 
saying that a group $x$ of Wick contractions is equivalent 
to another group $y$ (and writing $x\sim y$)  
if $x$ is obtained from $y$ by repetitive use of the relations 
\eqref{symmetry_Y} and \eqref{symmetry_C}
and by permuting interaction vertices of the same type. 
We denote the equivalence class of $x$ by $[x]=\{ y \,|\, y\sim x \}$, 
and call $\gamma=[x]$ a Feynman diagram.
The perturbative expansion of the free energy 
is then given by a sum over connected Feynman diagrams $\gamma$, 
$F=\sum_\gamma F(\gamma)$\,.  
In a connected diagram, 
every index line makes a loop since all the indices are contracted.

%%%%%%%%%%%%%%%%%%%%%%%%%%%%%%%%%%%%%%%
\subsection{Restriction to tetrahedral decompositions}
%%%%%%%%%%%%%%%%%%%%%%%%%%%%%%%%%%%%%%%

\begin{figure}[htbp]
\begin{quote}
\begin{center}
 \includegraphics[height = 3.0cm]{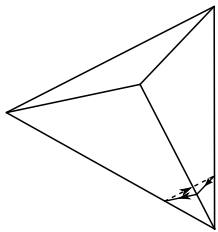}
 \caption{An index triangle made on three sides of thickened triangles, 
 which form a corner of a tetrahedron. 
 } 
\label{tetra_index_triangle}
\vspace{-3ex}
\end{center}
\end{quote}
\end{figure}

Although configurations generated in triangle-hinge models 
do not generally represent tetrahedral decompositions,
the set of Feynman diagrams can be reduced 
such that they represent only (and all of the) tetrahedral decompositions 
if we take a large $n$ limit   
with $n/\lambda$ and $n^2 \mu_k$ fixed \cite{Fukuma:2015xja}.%
\footnote{%=====
 The set of tetrahedral decomposition can be further restricted 
 so as to represent {\em manifolds} 
 by extending the algebra $\mathcal{A}$ as having a center 
 to count the number of vertices \cite{Fukuma:2015xja}. 
} %============= 
The point is the following. 
We have $\mathrm{tr} \,\omega^\ell$ 
when $\omega$ appears $\ell$ times in an index loop 
(see Fig.~\ref{thickened_triangle-hinge_index}), 
and thus $F(\gamma)$ is given by 
\begin{align}
 F(\gamma)=\frac{1}{S(\gamma)} 
 \Bigl(\frac{\lambda}{n^3}\Bigr)^{s_2(\gamma)} 
 \biggl[\,\prod_{k=1} (n^2 \mu_k)^{s_1^k (\gamma)} \biggr] 
 \, \prod_{\ell=1} [\mathrm{tr} \,\omega^\ell ]^{t_2^\ell(\gamma)} .
\label{general_f_gamma}
\end{align}
Here, $t_2^\ell(\gamma)$ denotes 
the numbers of index $\ell$-gons in diagram $\gamma$.%
\footnote{%=====
 An index loop is called an {\em index $\ell$-gon} 
 if it consists of $\ell$ intervals, 
 each living on a side of an intermediate triangle \cite{Fukuma:2015xja}. 
} %============= 
$s_2(\gamma)$ and $s_1^k (\gamma)$ denote 
the number of triangles and $k$-hinges in diagram $\gamma$,  respectively, 
and $S(\gamma)$ is the symmetry factor. 
Due to the definition of matrix $\omega$, 
we have 
\begin{align}
 \mathrm{tr}\,\omega^\ell = \left\{ \begin{array}{l}
 n \phantom{0}\hspace{1em} (\ell = 0 \mod 3) \\
 0 \phantom{n}\hspace{1em} (\ell \neq 0 \mod 3) .
 \end{array}\right.
\end{align} 
Thus, there can survive only the diagrams with $\ell$ a multiple of three 
($\ell\equiv 3\ell'$), 
and we can assume \eqref{general_f_gamma} to take the form 
\begin{align}
 F(\gamma) = \frac{1}{S(\gamma)} \lambda^{s_2(\gamma)}
 \biggl[\,\prod_{k\geq 1} (n^2 \mu_k)^{s_1^k (\gamma)} \biggr]
 \, n^{-3 s_2(\gamma) + \sum_{\ell'\geq 1} t_2^{3\ell'}(\gamma)} \,. 
\label{f_gamma}
\end{align}
One can show that 
only the index polygons with $\ell=3$ (i.e.\ $\ell'=1$) survive 
in the limit $n\to\infty$ with $n/\lambda$ and $n^2 \mu_k$ fixed 
\cite{Fukuma:2015xja}. 
We here give a proof in a form slightly different from the original one 
such that it can be applied to unoriented models.
We first note that the relation 
$
 \sum_{\ell' \geq 1} 3\ell' t_2^{3\ell'} (\gamma) = 6 s_2(\gamma)
$
holds because the left-hand side counts 
the number of $\omega$ in diagram $\gamma$ 
and each thickened triangle has six insertions of $\omega$. 
Then, if we introduce a nonnegative quantity 
$
 d(\gamma)\equiv \sum_{\ell'\geq 1} 3(\ell'-1) t^{3\ell'}_2(\gamma)\geq 0
$, 
we have the relation 
$
 d(\gamma) = 6 s_2(\gamma) -3 \sum_{\ell' \geq 1} t^{3\ell'}_2(\gamma). 
$
Thus, $\lambda^{s_2(\gamma)}$ 
% in $F(\gamma)$ [eq.~\eqref{f_gamma}] 
can be rewritten as 
\begin{align}
 \lambda^{s_2(\gamma)} =  \lambda^{-\frac13 d(\gamma)}
 \lambda^{3 s_2(\gamma) - \sum_{\ell' \geq 1} t^{3\ell'}_2(\gamma)} \,. 
\end{align}
Substituting this expression to \eqref{f_gamma}, $F(\gamma)$ is expressed as 
\begin{align}
F(\gamma) =  
\frac{1}{S(\gamma)} \lambda^{-\frac13 d(\gamma)}
 \biggl[\,\prod_{k\geq 1} (n^2 \mu_k)^{s_1^k (\gamma)} \biggr] 
 \, \Bigl(\frac{n}{\lambda}\Bigr)^{-3 s_2(\gamma) 
+ \sum_{\ell'\geq 1} t_2^{3\ell'}(\gamma)} \,.
\end{align}
Therefore,  in the limit $n\to \infty$ 
with $n/\lambda$ and $n^2 \mu_k$ fixed (and thus $\lambda\to \infty$), 
only the diagrams satisfying $d(\gamma)=0$ can give nonzero contributions 
to the free energy. 
Since $d(\gamma)=0$ means that all the index polygons in $\gamma$ are triangles, 
we conclude that 
the large $n$ limit reduces the set of diagrams 
so that all the index polygons are triangles. 
One can further prove 
that such diagrams represent tetrahedral decompositions \cite{Fukuma:2015xja}, 
as may be understood intuitively from the fact 
that if there is an index triangle, 
then sides of thickened triangles must be attached as in Fig.~\ref{tetra_index_triangle}.%
\footnote{%=====
 Note that there is $\omega$ at each corner of a side 
 of a thickened triangle.
} %============= 
%%%%%%%%%%%%%%%%%%%%%%%
%%%%%%%%%%%%%%%%%%%%%%%

We end this subsection with a comment. 
The above argument can also be applied 
to unoriented models to be defined in the next section. 
Namely, if a set of diagrams is reduced 
such that all the index polygons are triangles, 
then the diagrams represent tetrahedral decompositions 
even for unoriented models.

%%%%%%%%%%%%%%%%%%%%%%%%%%%%%%%%%%%%%%%
\subsection{Orientability}
\label{subsec_ori}
%%%%%%%%%%%%%%%%%%%%%%%%%%%%%%%%%%%%%%%

It is pointed out in \cite{Fukuma:2015xja} 
that all the tetrahedral decompositions 
generated by the action \eqref{oriented_THaction} are orientable. 
We here give a detailed proof of this statement, 
by clarifying the definition of orientation 
for Feynman diagrams in a triangle-hinge model. 

\begin{figure}[htbp]
\begin{quote}
\begin{center}
 \includegraphics[height = 4cm]{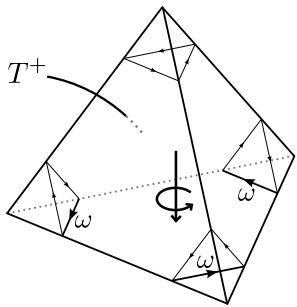}
 \caption{A positively oriented tetrahedra $T^+$, 
 corresponding to a positive embedding of $T$ in $E^3$. 
 % There is one index triangle at each corner. 
 }
\label{Fig1}
\vspace{-3ex}
\end{center}
\end{quote}
\end{figure}

We first recall that a thickened triangle has two triangular sides, 
on each of which directed index lines are drawn 
[see Fig.~\ref{thickened_vertices} (a)]. 
Given a tetrahedron $T$ formed by four triangular sides 
(each coming from a thickened triangle), 
we embed it to a three-dimensional Euclidean space $E^3$ 
as a regular tetrahedron of unit volume. 
Note that there can be two embeddings $f^+$ and $f^-$  
(up to rotations and translations in $E^3$), 
depending on whether the directions of index lines 
are counterclockwise or clockwise 
when seen from the center of the embedded tetrahedron 
(see Fig.~\ref{Fig1}).%
\footnote{%=====
 Note that if the directions of index lines are counterclockwise 
 for one side, 
 they are also counterclockwise for the other three sides 
 because index lines are connected 
 in such a way that the direction is preserved. 
} %============= 
%%%%%%%%%%%%%%%%%%%%%%%
%%%%%%%%%%%%%%%%%%%%%%%
We say the former embedding to be {\em positive} 
and the latter {\em negative}. 
We then define an {\em oriented tetrahedron} $T^\pm$ 
to be the pair of tetrahedron and embedding, 
$T^\pm \equiv (T,f^\pm)$.

\begin{figure}[htbp]
\begin{quote}
\begin{center}
 \includegraphics[height = 3.5cm]{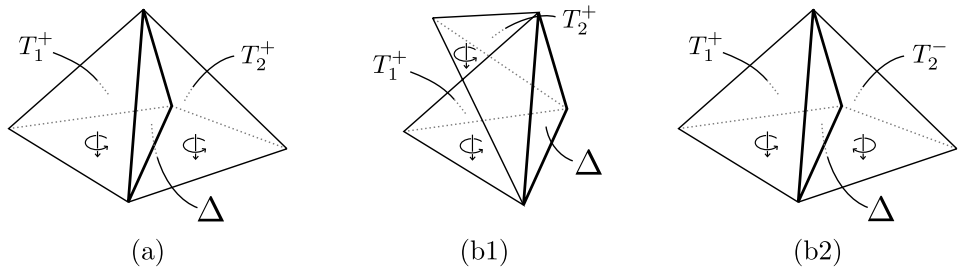}
 \caption{
 Two tetrahedra $T_1$ and $T_2$  
 glued at triangle $\Delta$.
 (a): both tetrahedra are positively oriented, 
 and the orientation is preserved 
 because they are in opposite positions with respect to $\Delta$. 
 (b1): both are positively oriented, 
 but the orientation is not preserved 
 because they are in the same position with respect to $\Delta$. 
 (b2): they are in opposite positions 
 but are differently oriented. 
 }
\label{Fig_tetra_embed}
\vspace{-3ex}
\end{center}
\end{quote}
\end{figure}

When two positively oriented tetrahedra $T_1^+$ and $T_2^+$ 
are glued at a triangle $\Delta$, 
we say that the orientation is preserved  
if the two positive embeddings $f_1^+$ and $f_2^+$ 
can be extended 
(with the use of rotations and translations)
to a common embedding $f$ of $T^+_1\cup T^+_2$  
such that the images of two tetrahedra are in opposite positions 
with respect to the intermediate triangle $\Delta$ 
(see Fig.~\ref{Fig_tetra_embed}). 
%%%%%%%%%%%%%%%%%%%%%%%
%%%%%%%%%%%%%%%%%%%%%%%
We then say that 
a tetrahedral decomposition $\Gamma$ is {\em orientable} 
if the orientation is preserved 
for any two adjacent tetrahedra of positive orientation.

The above orientability condition  
actually holds for tetrahedral decompositions 
% obtained from the triangle-hinge models 
discussed in the previous subsection. 
In fact, the index lines on the two sides of a thickened triangle 
are drawn in opposite directions 
as in Fig.~\ref{thickened_vertices} (a),  
and thus, for any two adjacent tetrahedra 
there always exists a natural extension of their positive embeddings 
such that the images of two tetrahedra 
are in opposite positions with respect to the triangle. 
Since it holds for every two adjacent tetrahedra, 
we conclude that all the tetrahedral decompositions are orientable.

%%%%%%%%%%%%%%%%%%%%%%%%%%%%%%%%%%%%%%%%%%%%%
%%%%%%%%%%%%%%%%%%%%%%%%%%%%%%%%%%%%%%%%%%%%%
\section{Unoriented membrane theories}
\label{sec_unorient}
%%%%%%%%%%%%%%%%%%%%%%%%%%%%%%%%%%%%%%%%%%%%%
%%%%%%%%%%%%%%%%%%%%%%%%%%%%%%%%%%%%%%%%%%%%%

In this section, 
we define unoriented membrane theories 
in terms of tetrahedral decompositions. 
A realization of unoriented membrane theories 
within the framework of triangle-hinge models 
will be given in the next section.

%%%%%%%%%%%%%%%%%%%%%%%%%%%%%%%%%%%%%%%%%%%%%
\subsection{Matrix models for unoriented strings}
%%%%%%%%%%%%%%%%%%%%%%%%%%%%%%%%%%%%%%%%%%%%%

As a warm-up before discussing unoriented membrane theories,  
we review the definition of unoriented string theories 
and how some of them are realized 
in terms of real symmetric matrix models.  

We first recall that 
an {\em oriented} open string is 
an oriented one-dimensional object with two ends.
If we forget about the target-space degrees of freedom,  
% and Chan-Paton indices which may be attached to the ends, 
the scattering processes of oriented open strings 
are represented by Feynman diagrams of Hermitian matrix models: 
\begin{align}
 S[M] = \mathrm{tr}
 \Bigl(\frac{1}{2} \,M^2 - \frac{\lambda}{3}\, M^3 \Bigr),
\label{mm_action}
\end{align}
where $M = (M_{ij}) = M^\dag$ is an $N \times N$ Hermitian matrix. 
In fact, the propagator and the interaction vertex 
are expressed as%
\footnote{%=====
 Note that 
 in $\mathrm{tr}(M^3) = C^{ijklmn} M_{ij}M_{kl} M_{mn}$, 
 only such components of $C^{ijklmn}$ survive 
 that are totally symmetric 
 under the permutation of three pairs of indices, 
 $(ij)$, $(kl)$, $(mn)$. 
 However, 
 when we write $C^{ijklmn} =\delta^{jk}\delta^{lm} \delta^{ni}$\,,
 we intensionally think that $C^{ijklmn}$ 
 are only cyclically symmetric for three pairs of indices,  
 and distinguish two diagrams, 
 one coming from $C^{ijklmn}$ 
 and the other from $C^{klijmn}$. 
 Of course, 
 by summing two ways of Wick contractions 
 in calculating the free energy, 
 the two diagrams appear in a combined way 
 and $C^{ijklmn} $ will be automatically symmetrized. 
 This ``trick'' enables us to identify a Feynman diagram 
 with a triangulated surface, 
 and is widely and implicitly adopted in the study of matrix models. 
} %============= 
\begin{align}
 {\rm propagator}&:\ \begin{array}{l} 
 \includegraphics[height = 0.9cm]{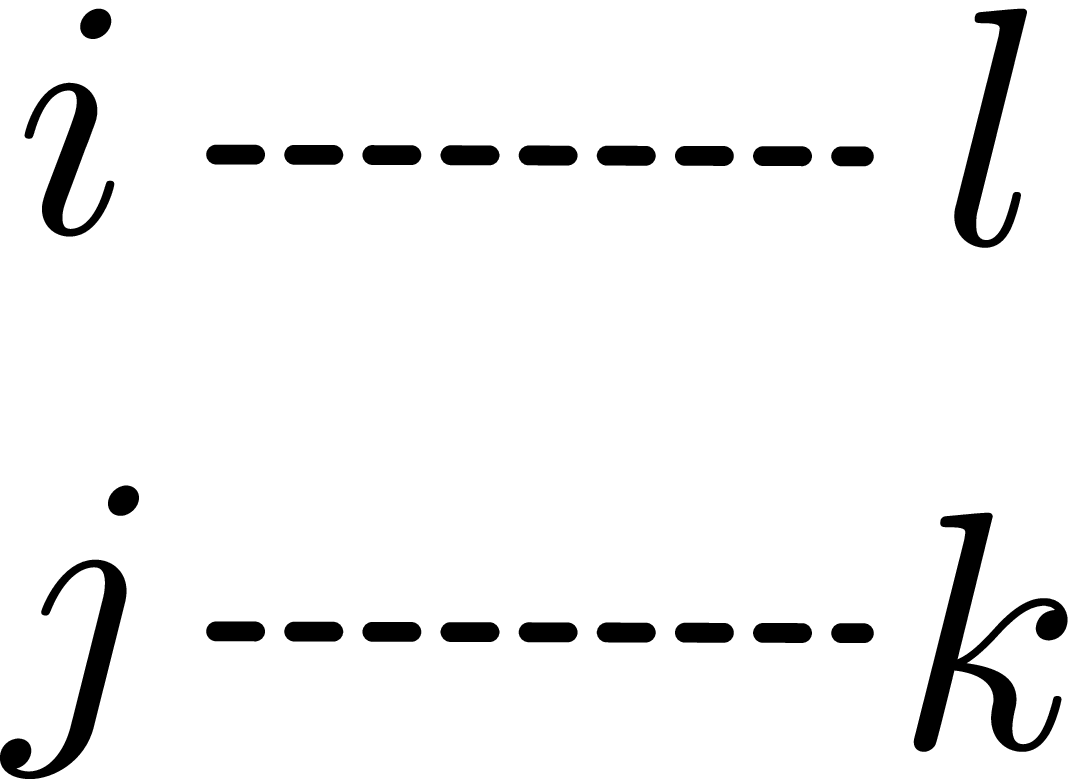} 
 \end{array} 
 \sim \delta_{il} \delta_{jk} \,\,
 \bigl( = \contraction{}{M}{_{ij}}{M} M_{ij}M_{kl} \bigr) ,
\label{mm_propagator} 
\\
 {\rm interaction}&:\ \begin{array}{l} 
 \includegraphics[width = 2.0cm]{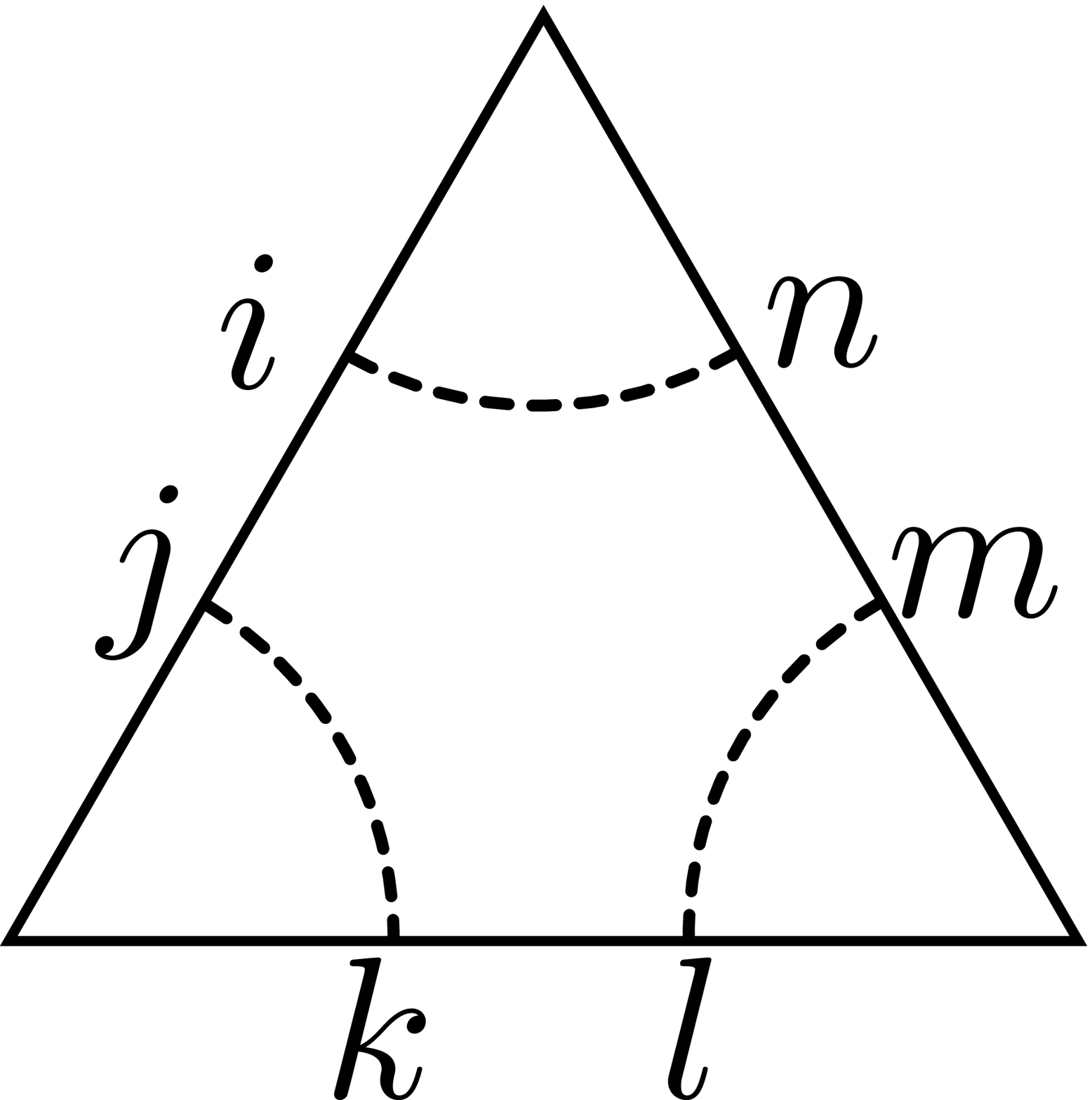}
 \end{array}
 \sim \lambda \, \delta^{jk}\,\delta^{lm}\,\delta^{ni} \,. 
\label{mm_triangle}
\end{align} 
Each Feynman diagram can also be thought of 
as a triangular decomposition 
of an orientable two-dimensional surface 
by representing it with the dual diagram.

We now introduce a transformation $\Omega$ 
which acts on one-string states 
and inverts the worldsheet parity (the orientation of string). 
{\em Unoriented} open string theories are then defined as theories 
where the transformation $\Omega$ is gauged 
(see, e.g., \cite{Polchinski:1998rq}). 
Namely, we demand that every propagator in the open-string channel 
be invariant under the action of $\Omega$\,. 
This is realized by inserting the projector $(1+\Omega)/2$ 
to every propagator. 
If we do not change the form of interaction, 
the Feynman rules are then expressed as follows
(we have rescaled the projector for later convenience): 
\begin{align}
 {\rm propagator}&:\ \begin{array}{l} 
 \includegraphics[height = 0.9cm]{mm_propagator.eps} 
 \end{array} 
 + \begin{array}{l} 
 \includegraphics[height = 0.9cm]{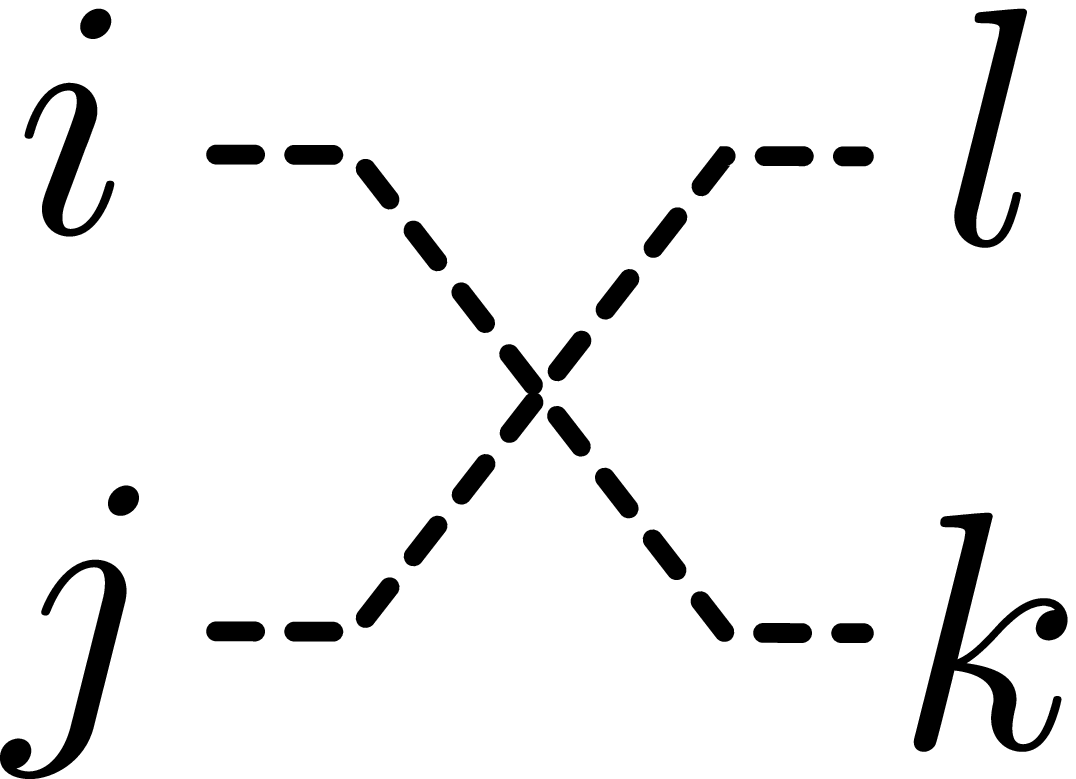}
 \end{array}  
 \sim \delta_{il} \delta_{jk} + \delta_{ik} \delta_{jl}\,\, 
 \bigl( = \contraction{}{X}{_{ij}}{X} X_{ij}X_{kl} \bigr) ,
\label{smm_propagator} 
\\
 {\rm interaction}&:\ \begin{array}{l} 
 \includegraphics[width = 2.0cm]{mm_tri.eps}
 \end{array}
 \sim \lambda \, \delta^{jk}\,\delta^{lm}\,\delta^{ni} . 
\label{smm_triangle} 
\end{align}
It is easy to see that 
the above Feynman rules are obtained 
from a real symmetric matrix model: 
\begin{align}
 S = \mathrm{tr}
 \Bigl(\frac{1}{4}\,X^2 - \frac{\lambda}{6}\, X^3 \Bigr) , 
\label{smm_action}
\end{align}
where $X=(X_{ij})=X^T$ is an $N\times N$ real symmetric matrix.

\begin{figure}[htbp]
\begin{quote}
\begin{center}
 \includegraphics[height = 5cm]{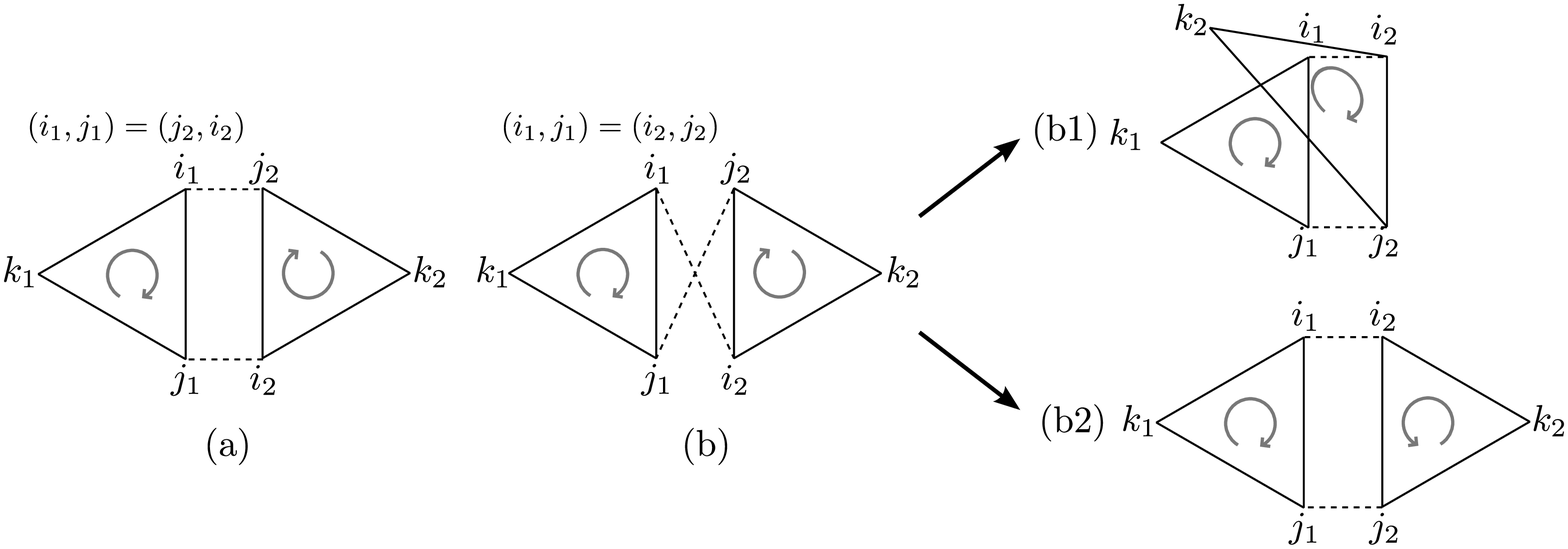}
 \caption{
 Two ways to identify edges of triangles of positive orientation. 
 The orientation is preserved for 
 (a): $(i_1,j_1) = (j_2,i_2)$, 
 but is not for 
 (b): $(i_1,j_1) = (i_2,j_2)$. 
 The local two-dimensional orientations of triangles 
 induce one-dimensional orientations of the edges to be identified. 
 The identification (b) can also be expressed as (b1) or (b2).  
 The expression (b2) is necessarily accompanied 
 by the flip of the right triangle, 
 which means that the local two-dimensional orientation 
 is not preserved 
 when one  moves from the left triangle to the right triangle 
 across the identified edge. 
 } 
\label{Fig_tri_propagator}
\vspace{-3ex}
\end{center}
\end{quote}
\end{figure}

A Feynman diagram for the above unoriented string theory 
can also be represented 
as a collection of triangles glued together along two-hinges. 
In fact, if we express the vertex $C^{ijklmn}$ 
(only with a cyclic symmetry) by an oriented triangle, 
the two contractions in \eqref{smm_propagator} 
can be illustrated as in Fig.~\ref{Fig_tri_propagator}. 
%%%%%%%%%%%%%%%%%%%%%%%
%%%%%%%%%%%%%%%%%%%%%%% 
The first contraction leads to a gluing of two oriented triangles 
with the orientation being preserved, 
while the second contraction to a gluing 
for which the orientation is not preserved.

%%%%%%%%%%%%%%%%%%%%%%%%%%%%%%%%%%%%%%%%%%%%%
\subsection{Unoriented membrane theories}
%%%%%%%%%%%%%%%%%%%%%%%%%%%%%%%%%%%%%%%%%%%%%

In the previous subsection, 
we have seen that unoriented string theories are obtained 
from oriented theories 
by gauging the worldsheet parity transformation $\Omega$. 
We now apply the same prescription 
to membrane theories 
in order to define {\em unoriented membrane theories}; 
We first prepare oriented models 
and introduce the worldvolume parity transformation $\Omega$ 
that inverts the orientation of open membrane, 
and then gauge the transformation $\Omega$ 
by inserting $(1+\Omega)/2$ to every propagator of open membrane   
in the original oriented models. 
In the rest of this paper, 
we assume that worldvolumes in oriented models are already represented 
as tetrahedral decompositions.

%%%%%%%%%%%%%%%%%%%%%%%%%%%%%%%%%%%%%%%%%%%%%
\subsubsection{Open membranes of disk topology as fundamental objects}

We first argue that the worldvolume dynamics of 
oriented {\em closed} membranes of various topologies 
can also be regarded as that of 
oriented {\em open membranes of disk topology}. 
In fact, tetrahedra in a tetrahedral decomposition 
can be thought of as interaction vertices 
that are connected with propagators of membrane 
of disk topology (i.e.\ triangles). 
One thus may say that a worldvolume theory of closed membranes 
of arbitrary topologies has a dual picture  
where open membranes of disk topology play fundamental roles, 
despite the fact that open membranes can have topologies other than disk 
(such as disks with handles).

%%%%%%%%%%%%%%%%%%%%%%%%%%%%%%%%%%%%%%%%%%%%%
\subsubsection{Fundamental triplets for oriented membranes}

\begin{figure}[htbp]
\begin{quote}
\begin{center}
 \includegraphics[height = 5cm]{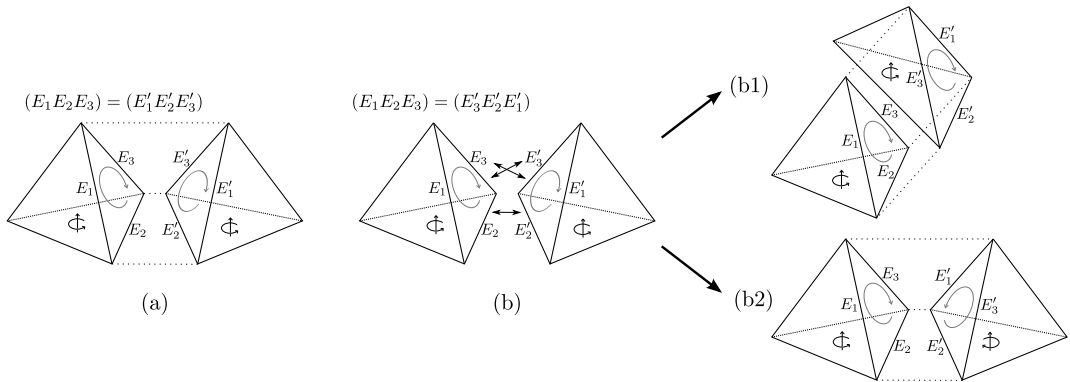}
 \caption{
 Two ways to identify triangles of tetrahedra of positive orientation. 
 The orientation is preserved for 
 (a): $(E_1 E_2 E_3)=(E_1^\prime E_2^\prime E_3^\prime)$, 
 but is not for 
 (b): $(E_1 E_2 E_3)=(E_3^\prime E_2^\prime E_1^\prime)$. 
 The local three-dimensional orientations of tetrahedra 
 induce two-dimensional orientations of the triangles to be identified. 
 The identification (b) can also be expressed as (b1) or (b2). 
 The expression (b2) is necessarily accompanied 
 by the orientation change of the right tetrahedron, 
 which means that the local three-dimensional orientation 
 is not preserved when 
 one moves from the left tetrahedron to the right tetrahedron 
 across the identified triangle.  
 }
\label{Fig_tetra_id}
\vspace{-3ex}
\end{center}
\end{quote}
\end{figure}

Given an oriented model, 
we focus on two adjacent, positively oriented tetrahedra 
$T^+_1$ and $T^+_2$ 
in a tetrahedral decomposition $\Gamma$, 
where $T^+_1$ and $T^+_2$ are glued 
by identifying a triangle $\Delta_1$ in $T^+_1$ 
with a triangle $\Delta_2$ in $T^+_2$ 
(the resulting identified triangle will be denoted by $\Delta$). 
Note that the orientation of a tetrahedron naturally induces 
the positive orientation for four triangles belonging to the tetrahedron, 
and we represent them by arrows as in Fig.~\ref{Fig_tetra_id} (a). 
%%%%%%%%%%%%%%%%%%%%%%%
%%%%%%%%%%%%%%%%%%%%%%% 
We express the identification of edges at $\Delta$ as 
\begin{align}
(E_1 E_2 E_3)=(E_1^\prime E_2^\prime E_3^\prime) \,,  
\label{gamma_identification}
\end{align}
where $E_1, E_2, E_3$ (or $E_1^\prime, E_2^\prime, E_3^\prime$) 
are the edges of $\Delta_1$ (or $\Delta_2$).  
Note that the orientations of $\Delta_1$ and $\Delta_2$ must be opposite 
in order to form an oriented tetrahedral decomposition, 
and thus the three-dimensional orientation is preserved 
when one moves from the inside of $T_1$
to that of $T_2$ through the identified triangle $\Delta$ 
[Fig.~\ref{Fig_tetra_id} (a)]. 
Three-dimensional orientation is also preserved 
for the two other tetrahedral decompositions 
that are obtained from \eqref{gamma_identification} 
by cyclically permuting the edges $(E'_1,E'_2,E'_3)$. 
We denote by $\Gamma_1(=\Gamma)$, $\Gamma_2$, $\Gamma_3$, respectively,  
the tetrahedral decompositions 
corresponding to the three edge-identifications 
that preserve the orientation, 
\begin{align}
 \Gamma_1:\,(E_1 E_2 E_3)=(E_1^\prime E_2^\prime E_3^\prime),\quad
 \Gamma_2:\,(E_1 E_2 E_3)=(E_2^\prime E_3^\prime E_1^\prime),\quad
 \Gamma_3:\,(E_1 E_2 E_3)=(E_3^\prime E_1^\prime E_2^\prime)\,.  
\label{++}
\end{align}
We will call $(\Gamma_1, \Gamma_2, \Gamma_3)$ 
the {\em fundamental triplet associated with triangle $\Delta$}.

%%%%%%%%%%%%%%%%%%%%%%%%%%%%%%%%%%%%%%%%%%%%%
\subsubsection{Definition of unoriented membrane theories}

In addition to the edge-identifications \eqref{++} 
[leading to the fundamental triplet $(\Gamma_1,\Gamma_2,\Gamma_3)$]\,, 
we introduce another triplet $(\tilde\Gamma_1,\tilde\Gamma_2,\tilde\Gamma_3)$ 
that are obtained, respectively, by the following edge-identifications 
at the same triangle $\Delta$\,: 
\begin{align}
 \tilde\Gamma_1:\,(E_1 E_2 E_3)=(E_3^\prime E_2^\prime E_1^\prime),\quad
 \tilde\Gamma_2:\,(E_1 E_2 E_3)=(E_1^\prime E_3^\prime E_2^\prime),\quad
 \tilde\Gamma_3:\,(E_1 E_2 E_3)=(E_2^\prime E_1^\prime E_3^\prime)\,. 
\label{+-}
\end{align}
Note that, in contrast to \eqref{++}, 
three-dimensional orientation is not preserved across $\Delta$  
[see Fig.~\ref{Fig_tetra_id} (b)]. 
We introduce a transformation $\Omega$ 
that interchanges two triplets 
$(\Gamma_1,\Gamma_2,\Gamma_3)$ 
and $(\tilde\Gamma_1,\tilde\Gamma_2,\tilde\Gamma_3)$, 
and define {\em unoriented membrane theories} 
to be those that are obtained from the oriented theories 
by acting the projection operator $(1+\Omega)/2$ 
on every triangle. 
We will call the set 
$(\Gamma_1,\Gamma_2,\Gamma_3,\tilde\Gamma_1,\tilde\Gamma_2,\tilde\Gamma_3)$ 
the {\em fundamental sextet associated with triangle $\Delta$}\,. 
So far we have assumed that 
the tetrahedral decomposition $\Gamma_1$ is orientable, 
but one can easily see that $\Gamma_1$ is not necessarily orientable 
for the above definition of a sextet to make sense 
because we focus only on local configurations around triangle $\Delta$. 
In the rest of paper, 
we understand that the domain of definition for $\Omega$ 
is extended so as to include nonorientable tetrahedral decompositions.

Note that 
each sextet $(\Gamma_1,\ldots,\tilde\Gamma_3)$ 
consists of both manifolds and nonmanifolds, 
unlike the two-dimensional cases 
where $\Omega$ always relates a manifold to another manifold.  
In fact, suppose that a tetrahedral decomposition $\Gamma_1$ 
represents a three-dimensional manifold. 
Then, the change of the edge-identification at $\Delta$ 
from $(E_1 E_2 E_3)=(E_1^\prime E_2^\prime E_3^\prime)$ 
to $(E_1 E_2 E_3)=(E_3^\prime E_2^\prime E_1^\prime)$ 
gives rise to a singularity
at the midpoint of edge $E_2=E_2^\prime$ in $\tilde\Gamma_1$ 
around which we cannot define a local orientation. 
The appearance of singularity will be demonstrated explicitly 
when we consider an example in subsection \ref{subsec_example}.

%%%%%%%%%%%%%%%%%%%%%%%%%%%%%%%%%%%%%%%%%%%%%
%%%%%%%%%%%%%%%%%%%%%%%%%%%%%%%%%%%%%%%%%%%%%
\section{Triangle-hinge models for unoriented membranes}
\label{sec_unorient_th}
%%%%%%%%%%%%%%%%%%%%%%%%%%%%%%%%%%%%%%%%%%%%%
%%%%%%%%%%%%%%%%%%%%%%%%%%%%%%%%%%%%%%%%%%%%%

%%%%%%%%%%%%%%%%%%%%%%%%%%%%%%%%%%%%%%%%%%%%%
\subsection{Action and Feynman rules}
\label{subsec_action_tet}
%%%%%%%%%%%%%%%%%%%%%%%%%%%%%%%%%%%%%%%%%%%%%

In this section, 
we realize unoriented membrane theories as triangle-hinge models. 
We show that they are obtained 
simply by replacing $C=C_+$ in the original oriented models \eqref{oriented_THaction} 
with $C=C_+ + C_-$: 
\begin{align}
 S &=\frac{1}{2}\,[AB]
 - \frac{\lambda}{6}\, \bigl([C_{+}AAA] + [C_{-}AAA]\bigr) 
 - \sum_k \frac{\mu_k}{2k} \,[Y_k \underbrace{B \cdots B}_k] 
\nn\\
 &\equiv\frac{1}{2}A_{abcd}B_{abcd} 
\nn\\ 
 &~~ -\frac{\lambda}{6n^3}\,
 (\omega^{d_1 a_2}\omega^{d_2 a_3}\omega^{d_3 a_1}
 + \omega^{d_3 a_2}\omega^{d_2 a_1}\omega^{d_1 a_3}) \,
 \omega^{b_3 c_2}\omega^{b_2 c_1}\omega^{b_1 c_3}\, 
 A_{a_1 b_1 c_1 d_1} A_{a_2 b_2 c_2 d_2} A_{a_3 b_3 c_3 d_3} 
\nn\\ 
 &~~ - \sum_k \frac{n^2 \mu_k}{2k} B_{a_1 a_2 b_2 b_1} 
 \cdots B_{a_{k-1} a_k b_k b_{k-1}} B_{a_k a_1 b_1 b_k} \,.
\label{unoriented_THaction}
\end{align}
Here, 
$C_{+}$ is again given by eq.~\eqref{cplus}, 
and $C_{-}$ by 
\begin{align}
 C_{-}^{a_1 b_1 c_1 d_1 a_2 b_2 c_2 d_2 a_3 b_3 c_3 d_3} 
 \equiv \frac{1}{n^3}\, 
\omega^{d_3 a_2}\omega^{d_2 a_1}\omega^{d_1 a_3}
 \omega^{b_3 c_2} \omega^{b_2 c_1} \omega^{b_1 c_3} \,. 
\label{cminus}
\end{align}

\begin{figure}[htbp]
\begin{quote}
\begin{center}
 \includegraphics[height = 3.0cm]{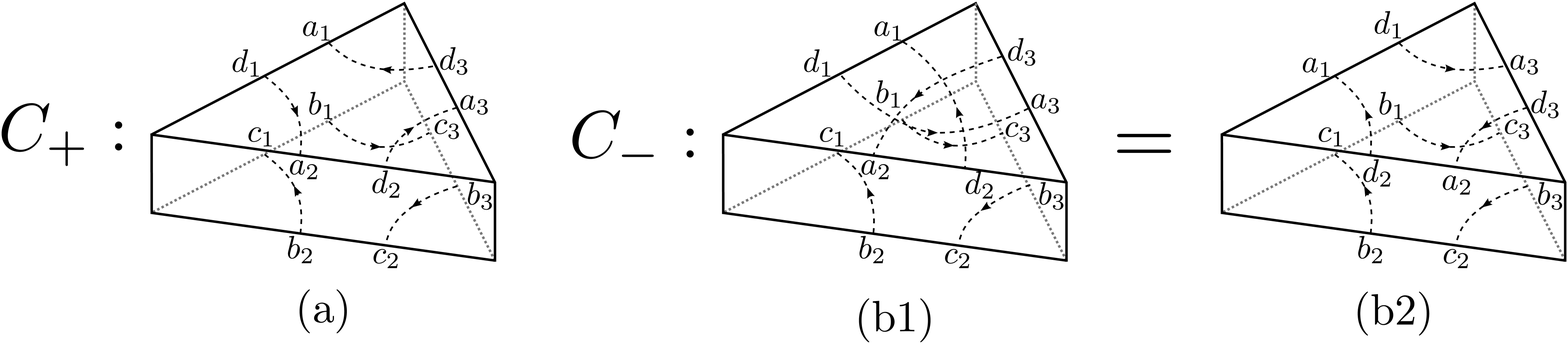}
 \caption{Interaction vertices corresponding to 
 (a):\,$[C_{+}AAA]$ and (b):\,$[C_{-}AAA]$. 
 If $C_{+}$ represents triangle-identifications 
 which preserve three-dimensional local orientation,  
 $C_{-}$ represents triangle-identifications 
 which do not preserve orientation. 
}
\label{Fig_tri_+-}
\vspace{-3ex}
\end{center}
\end{quote}
\end{figure}

We first note that 
the interaction vertices corresponding to $[C_{+}AAA]$ and $[C_{-}AAA]$ 
can be expressed by thickened triangles with directed index lines 
as in Fig.~\ref{Fig_tri_+-}. 
%%%%%%%%%%%%%%%%%%%%%%%
%%%%%%%%%%%%%%%%%%%%%%%
Contractions using $[C_+ AAA]$ yield 
the identification of a triangle belonging to a tetrahedron 
with another triangle belonging to an adjacent tetrahedron 
so that the orientation is preserved [see Fig.~\ref{Fig_tri_+-} (a)]. 
The two positively oriented tetrahedra thus reside in opposite positions 
with respect to the thickened triangle, 
and the edges $(a_1,d_1)$, $(a_2,d_2)$, $(a_3,d_3)$ 
will be identified 
with the edges $(b_1,c_1)$, $(b_2,c_2)$, $(b_3,c_3)$, respectively, 
when we deflate the triangle to get a tetrahedral decomposition.  
On the contrary, 
contractions using $[C_- AAA]$ yield 
an identification of triangles 
where the orientation is not preserved 
[see Fig.~\ref{Fig_tri_+-} (b1,b2)]. 
In fact, the indices of $C_-$ [see \eqref{cminus}] 
can be expressed as Fig.~\ref{Fig_tri_+-} (b1) or (b2). 
We use the expression (b1) 
in a Feynman diagram where the triangle is connected to hinges, 
but we exploit the other expression (b2) 
when the thickened triangle is interpreted 
as representing two triangles to be identified 
in gluing two tetrahedra of positive orientation. 
Then, the edges $(a_1,d_1)$, $(a_2,d_2)$, $(a_3,d_3)$ 
will be identified 
with the edges $(c_1,b_1)$, $(c_2,b_2)$, $(c_3,b_3)$, respectively, 
when we deflate triangles to get a tetrahedral decomposition.  
It is easy to see that the two positively oriented tetrahedra 
are now in the same position with respect to the triangle 
and thus will take a configuration of Fig.~\ref{Fig_tetra_embed} (b1) 
after the triangle is deflated. 
This means that the orientation is not preserved 
for this gluing of tetrahedra.

Note that 
the direction of arrows on index lines is still preserved 
for diagrams using $C_{-}$. 
Thus, taking the same large $n$ limit as in the oriented models, 
we can reduce the set of diagrams 
such that all their index polygons are triangles,%
\footnote{%=====
 As in the original models, 
 a tetrahedron has one index triangle at each corner 
 (see Fig.~\ref{tetra_index_triangle}).
} %=============
and can conclude that they represent tetrahedral decompositions.

%%%%%%%%%%%%%%%%%%%%%%%%%%%%%%%%%%%%%%%%%%%%%
\subsection{Wick contractions corresponding to the fundamental sextet}
%%%%%%%%%%%%%%%%%%%%%%%%%%%%%%%%%%%%%%%%%%%%%

Recall that for each triangle $\Delta$ 
in a tetrahedral decomposition $\Gamma=\Gamma_1$,  
we have the fundamental sextet of tetrahedral decompositions, 
$(\Gamma_1,\Gamma_2,\Gamma_3,
\tilde\Gamma_1,\tilde\Gamma_2,\tilde\Gamma_3)$, 
which close among themselves under the action of $\Omega$\,. 
In this subsection, 
we write down the corresponding sextet 
$(\gamma_1,\gamma_2,\gamma_3,
\tilde\gamma_1,\tilde\gamma_2,\tilde\gamma_3)$ 
in unoriented triangle-hinge models.

We first note that, 
while the {\em total} number of triangles 
(as well as that of tetrahedra) 
is the same among the sextet $(\Gamma_1,\ldots,\tilde\Gamma_3)$, 
this is not the case for those numbers 
around each edge of triangle $\Delta$\,. 
For example, let us consider the case 
where the three edges of $\Delta$ in $\Gamma_1$ 
[denoted by $E_1(=E_1^\prime),E_2(=E_2^\prime),E_3(=E_3^\prime)$] 
are connected to three different hinges. 
If we change the identification at $\Delta$ 
from $(E_1 E_2 E_3)=(E_1^\prime E_2^\prime E_3^\prime)$ 
to $(E_1 E_2 E_3)=(E_2^\prime E_3^\prime E_1^\prime)$ 
to obtain $\Gamma_2$, 
all the three edges $E_1, E_2, E_3$ must be the same 
due to triangle-identifications at other triangles.%
\footnote{%=====
 Since we assume that $\Gamma$ is a tetrahedral decomposition 
 without boundaries, 
 other triangle-identifications in $\Gamma$ ensure 
 the edge-identifications $E_1=E_1^\prime$, $E_2=E_2^\prime$ 
 and $E_3=E_3^\prime$. 
 See discussions below \eqref{Gamma_1c} for more details. 
}  %===========
Therefore, 
Feynman diagrams $\gamma_1$ and $\gamma_2$ in a triangle-hinge model 
must have different numbers and different types of hinges 
if they correspond to $\Gamma_1$ and $\Gamma_2$, respectively.  
This means that the constructions of sextets 
are not so straightforward in triangle-hinge models 
compared to other models (such as tensor models).

Let us make the above consideration to a more concrete form, 
considering a triangle $\Delta$ in a tetrahedral decomposition $\Gamma_1$, 
at which two positively oriented tetrahedra glued 
with the orientation being preserved. 
We first note that 
there are the following three cases 
for the three edges $I$, $J$, $K$ of triangle $\Delta$: 
\begin{align}
 &\mbox{(1)~} 
 \mbox{Three edges $I$, $J$, $K$ are connected to three different hinges.}
 \hspace{30mm}
\nn\\
 &\mbox{(2)~} 
 \mbox{Two and only two of them are connected to the same hinge.}
\nn\\
 &\mbox{(3)~} \mbox{All of them are connected to the same hinge.}
\label{edge-types}
\end{align}
We suppose that $\Gamma_1$ is of the type (1) at $\Delta$, 
and that edges $I$, $J$, $K$ 
are connected to $(p+1)$-, $(q+1)$-, $(r+1)$-hinges, respectively.
Including $p$ other edges connected to the $(p+1)$-hinge,  
we label the edges around the $(p+1)$-hinge 
as $[I,I_1,\ldots,I_p]$ in a cyclic order. 
Here, we define the cyclic ordering of edges around a $k$-hinge 
as follows (see Fig.~\ref{Fig_cyclic_ordering}): 
%%%%%%%%%%%%%%%%%%%%%%%
\begin{figure}[htbp]
\begin{quote}
\begin{center}
 \includegraphics[height = 4cm]{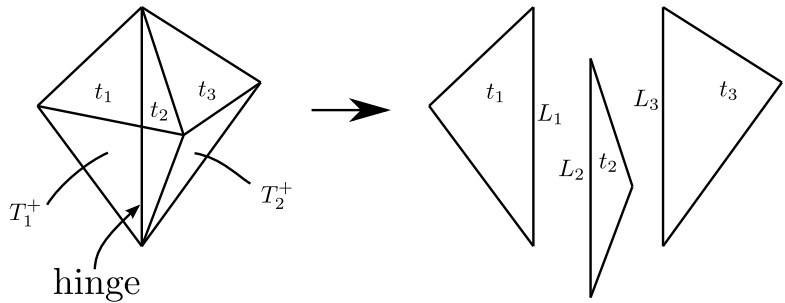}
 \caption{
 Labeling of edges around a $k$-hinge.  
 }
\label{Fig_cyclic_ordering}
\vspace{-3ex}
\end{center}
\end{quote}
\end{figure}
%%%%%%%%%%%%%%%%%%%%%%% 
We first pick up two neighboring triangles $t_1$ and $t_2$ 
belonging to the same tetrahedron $T^+_1$ of positive orientation, 
and label their edges connected to the hinge as $L_1$ and $L_2$, 
respectively. 
Next to $T^+_1$ 
there is another positively oriented tetrahedron $T^+_2$ 
determined by triangle $t_2$ and another triangle $t_3$ 
sharing the same hinge, 
and we label as $L_3$ the edge of $t_3$ that is connected to the hinge. 
Repeating this procedure, 
we obtain a sequence $[L_1,L_2,\ldots,L_k]$ around the $k$-hinge. 
Since another choice ($t_2,t_3$) is possible as the initial pair 
for the same configuration of edges around the hinge, 
we should regard the above sequence as being cyclically symmetric, 
$[L_1,L_2,\ldots,L_k]=[L_2,L_3,\ldots,L_k,L_1]$. 
Note that, 
if we take $(t_2,t_1)$ as the initial pair, 
the edges around the $k$-hinge will be represented 
as a sequence in reverse order, $[L_k,\ldots,L_2,L_1]$.

Labeling similarly the edges around the $(q+1)$- and $(r+1)$-hinges 
by $[J,J_1,\ldots,J_q]$ and $[K,K_1,\ldots,K_r]$, respectively, 
we have 
\begin{align}
 \Gamma_1:~
 & \mbox{edge $I$ is connected to a $(p+1)$-hinge as $[I,I_1,\ldots,I_p]$,}
\nn\\
 & \mbox{edge $J$ is connected to a $(q+1)$-hinge as $[J,J_1,\ldots,J_q]$,}
\nn\\
 & \mbox{edge $K$ is connected to a $(r+1)$-hinge as $[K,K_1,\ldots,K_r]$.} 
\label{Gamma_1}
\end{align}
Then, the remaining tetrahedral decompositions in the sextet 
have the following configurations: 
\begin{align}
 \Gamma_2:~
 & \mbox{three edges $I$, $J$, $K$ are connected to a single $(p+q+r+3)$-hinge}
\nn\\
 & \mbox{as $[I,I_1,\ldots,I_p,K,K_1,\ldots,K_r,J,J_1,\ldots,J_q]$. } 
\label{Gamma_2}\\
 \Gamma_3:~
 & \mbox{three edges $I$, $J$, $K$ are connected to a single $(p+q+r+3)$-hinge}
\nn\\
 & \mbox{as $[I,I_1,\ldots,I_p,J,J_1,\ldots,J_q,K,K_1,\ldots,K_r]$.} 
\label{Gamma_3}\\
 \tilde\Gamma_1:~
 & \mbox{two edges $I$, $K$ are connected to a $(p+r+2)$-hinge 
 as $[I,I_1,\ldots,I_p,K,K_1,\ldots,K_r]$,}
\nn\\
 & \mbox{and edge $J$ is connected to a $(q+1)$-hinge 
 as $[J,J_1,\ldots,J_q]$. } 
\label{tGamma_1}\\
 \tilde\Gamma_2:~
 & \mbox{two edges $J$, $K$ are connected to a $(q+r+2)$-hinge  
 as $[J,J_1,\ldots,J_q,K,K_1,\ldots,K_r]$, }
\nn\\
 & \mbox{and edge $I$ is connected to a $(p+1)$-hinge as $[I,I_1,\ldots,I_p]$. } 
\label{tGamma_2}\\
 \tilde\Gamma_3:~
 & \mbox{two edges $I$, $J$ are connected to a $(p+q+2)$-hinge  
 as $[I,I_1,\ldots,I_p,J,J_1,\ldots,J_q]$,}
\nn\\
 & \mbox{and edge $K$ is connected to a $(r+1)$-hinge 
 as $[K,K_1,\ldots,K_r]$.} 
\label{tGamma_3}
\end{align}

\begin{figure}[htbp]
\begin{quote}
\begin{center}
 \includegraphics[height = 3.5cm]{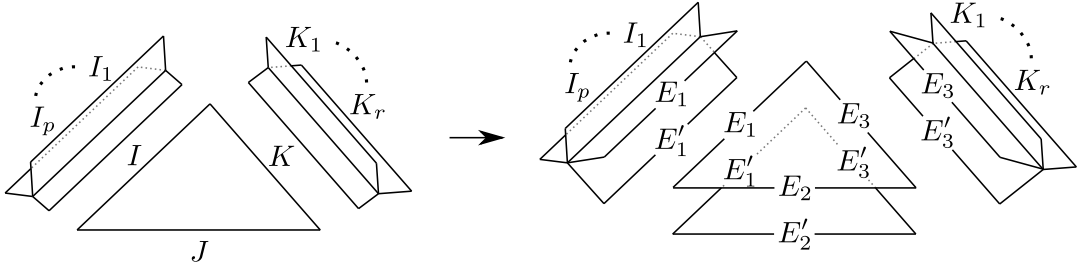}
 \caption{
 The splitting of $\Delta$ corresponding to $\Gamma_1$. 
 Edge $I$ becomes two edges $E_1$ and $E'_1$, 
 and edge $J$ (or $K$) becomes $E_2$ and $E'_2$ 
 (or to $E_3$ and $E'_3$). 
 The $(p+1)$-, $(q+1)$-, $(r+1)$-hinges accordingly become 
 $(p+2)$-, $(q+2)$-, $(r+2)$-hinges, respectively. 
 $\Gamma_1$ is restored by the edge-identification 
 $(E_1 E_2 E_3)=(E'_1 E'_2 E'_3)$ for the split triangles. 
 }
\label{Fig_tri_inflate}
\vspace{-3ex}
\end{center}
\end{quote}
\end{figure}

Equations \eqref{Gamma_1}--\eqref{tGamma_3} 
can be understood in the following way. 
We begin with \eqref{Gamma_1}, 
which is simplest and obvious.  
We first split the triangle $\Delta$ in $\Gamma_1$ 
to two triangles as in Fig.~\ref{Fig_tri_inflate} 
in order to realize the configuration 
before the edge-identification $(E_1 E_2 E_3)=(E'_1 E'_2 E'_3)$ is made.  
%%%%%%%%%%%%%%%%%%%%%%%
%%%%%%%%%%%%%%%%%%%%%%% 
This splitting is accompanied 
by that of edge $I$ to two edges $E_1$ and $E'_1$, 
and that of edge $J$ (or $K$) to $E_2$ and $E'_2$ 
(or to $E_3$ and $E'_3$). 
Accordingly, the $(p+1)$-, $(q+1)$- $(r+1)$-hinges are transformed 
to $(p+2)$-, $(q+2)$- $(r+2)$-hinges, respectively. 
Now we follow the sequence of the edges connected to each hinge  
in the other way around. 
If we start from the edge $E_1$ connected to the $(p+2)$-hinge, 
we then pass through the edges $I_1,\ldots,I_p$ 
following the original sequence $[I,I_1,\ldots,I_p]$, 
and reach the edge $E'_1$, 
which will be identified with the starting edge $E_1$ 
(i.e., $E'_1=E_1=I$)
under the edge-identification for $\Gamma_1$, 
$(E_1 E_2 E_3)=(E'_1 E'_2 E'_3)$. 
Let us write the total path schematically as a cycle, 
\begin{align}
 \bullet~
 E_1 \,\to\, I_1 \,\to\, \cdots \,\to\, I_p \,\to\, E'_1 = E_1. 
\label{Gamma_1a}
\end{align}
Similarly, if we start 
from the edge $E_2$ connected to the $(q+2)$-hinge 
or from the edge $E_3$ connected to the $(r+2)$-hinge,  
we then have the following paths: 
\begin{align}
 \bullet~&
 E_2 \,\to\, J_1 \,\to\, \cdots \,\to\, J_q \,\to\, E'_2 = E_2,
\label{Gamma_1b}
\\
 \bullet~&
 E_3 \,\to\, K_1 \,\to\, \cdots \,\to\, K_r \,\to\, E'_3 = E_3. 
\label{Gamma_1c}
\end{align}
Equations \eqref{Gamma_1a}--\eqref{Gamma_1c} 
are exactly what is expressed in \eqref{Gamma_1}.

\begin{figure}[htbp]
\begin{quote}
\begin{center}
 \includegraphics[height = 3.5cm]{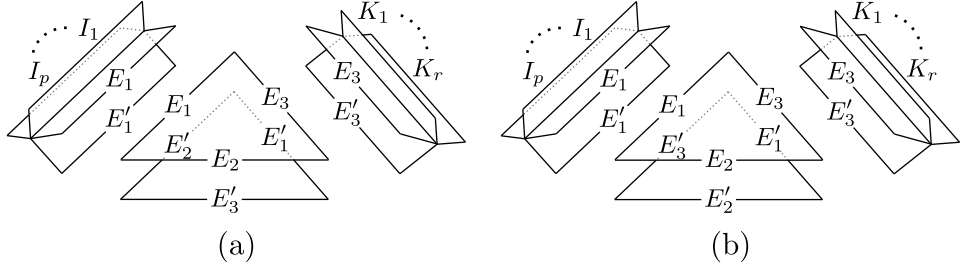}
 \caption{
 The splittings of $\Delta$ 
 corresponding to (a): $\Gamma_2$ and (b): $\tilde\Gamma_1$. 
 $\Gamma_2$ is obtained by the edge-identification 
 $(E_1 E_2 E_3)=(E'_2 E'_3 E'_1)$ for the split triangles, 
 and $\tilde\Gamma_1$ 
 by $(E_1 E_2 E_3)=(E'_3 E'_2 E'_1)$ for the split triangles. 
 }
\label{Fig_tri_inflate_other}
\vspace{-3ex}
\end{center}
\end{quote}
\end{figure}

Now we consider the tetrahedral decomposition $\Gamma_2$, 
which was obtained from $\Gamma_1$ 
by changing the edge-identification 
from $(E_1 E_2 E_3)=(E'_1 E'_2 E'_3)$ 
to $(E_1 E_2 E_3)=(E'_2 E'_3 E'_1)$ 
[see Fig.~\ref{Fig_tri_inflate_other} (a)]. 
%%%%%%%%%%%%%%%%%%%%%%%
%%%%%%%%%%%%%%%%%%%%%%% 
If we start from the edge $E_1$ connected to the $(p+2)$-hinge, 
we then again pass through the edges $I_1,\ldots,I_p$ 
and reach the edge $E'_1$. 
However, this is not the end of journey  
because $E'_1$ will be identified with $E_3$ in the edge-identification, 
and we need to follow another sequence of edges, 
$K_1,\ldots,K_r$, to reach $E'_3$. 
Since $E'_3$ will be identified with $E_2$, 
we need to continue the journey; 
we pass through the edges $J_1,\ldots,J_q$ to reach $E'_2$, 
which finally will agree with the starting edge $E_1$.  
The total path thus can be written as the following cycle: 
\begin{align}
 \bullet~
 &E_1 \,\to\, I_1 \,\to\, \cdots \,\to\, I_p \,\to\, E'_1 = E_3
 \,\to\, K_1 \,\to\, \cdots \,\to\, K_r
\nn\\
 &\,\to\, E'_3 = E_2 \,\to\, J_1 \,\to\, \cdots \,\to\, J_q 
 \,\to\, E'_2 = E_1.
\label{Gamma_2_sequence}
\end{align}
Since $E'_1=E_3=K$, $E'_3=E_2=J$ and $E'_2=E_1=I$ 
under the edge-identification, 
the path \eqref{Gamma_2_sequence} can be written as \eqref{Gamma_2}. 
Similarly, \eqref{Gamma_3} can be understood from the path:
\begin{align}
 \bullet~
 &E_1 \,\to\, I_1 \,\to\, \cdots \,\to\, I_p \,\to\, E'_1 = E_2
 \,\to\, J_1 \,\to\, \cdots \,\to\, J_q
\nn\\
 &\,\to\, E'_2 = E_3 \,\to\, K_1 \,\to\, \cdots \,\to\, K_r \,\to\, E'_3 = E_1.
\end{align}

Equation \eqref{tGamma_1} can be understood in a similar way, 
by recalling that $\tilde\Gamma_1$ is obtained from $\Gamma_1$ 
by changing the edge-identification 
from $(E_1 E_2 E_3)=(E'_1 E'_2 E'_3)$ 
to $(E_1 E_2 E_3)=(E'_3 E'_2 E'_1)$ 
[see Fig.~\ref{Fig_tri_inflate_other} (b)],  
which gives the following two disconnected cycles: 
\begin{align}
 \bullet~
 &E_1 \,\to\, I_1 \,\to\, \cdots \,\to\, I_p \,\to\, E'_1 = E_3
 \,\to\, K_1 \,\to\, \cdots\,\to\, K_r \,\to\, E'_3 = E_1,
\nn\\
 \bullet~
 &E_2 \,\to\, J_1 \,\to\, \cdots \,\to\, J_q \,\to\, E'_2 = E_2.
\end{align}
Similarly, the cycles for $\tilde\Gamma_2$ 
[edge-identification $(E_1 E_2 E_3)=(E'_1 E'_3 E'_2)$] 
are given by 
\begin{align}
 \bullet~
 &E_2 \,\to\, J_1 \,\to\, \cdots \,\to\, J_q \,\to\, E'_2 = E_3
 \,\to\, K_1 \,\to\, \cdots\,\to\, K_r \,\to\, E'_3 = E_2,
\nn\\
 \bullet~
 &E_1 \,\to\, I_1 \,\to\, \cdots \,\to\, I_p \,\to\, E'_1 = E_1 , 
\end{align}
and those for $\tilde\Gamma_3$ 
[edge-identification $(E_1 E_2 E_3)=(E'_2 E'_1 E'_3)$] 
are given by 
\begin{align}
 \bullet~
 &E_1 \,\to\, I_1 \,\to\, \cdots \,\to\, I_p \,\to\, E'_1 = E_2 
 \,\to\, J_1 \,\to\, \cdots \,\to\, J_q \,\to\, E'_2 = E_1,
\nn\\
 \bullet~
 &E_3 \,\to\, K_1 \,\to\, \cdots\,\to\, K_r \,\to\, E'_3 = E_3.
\end{align}

Now that we understand in detail the configurations of 
the fundamental sextet $(\Gamma_1,\ldots,\tilde\Gamma_3)$ 
associated with triangle $\Delta$, 
it is easy to translate \eqref{Gamma_1}--\eqref{tGamma_3} 
in terms of unoriented triangle-hinge models,  
and we obtain the sextet of Feynman diagrams 
($\gamma_1,\ldots,\tilde\gamma_3)$ 
as the following groups of Wick contractions:%
\footnote{%=====
 Due to the cyclic symmetry of $C_+$\,, 
 a group of wick contractions containing $[C_+ A_I A_J A_K]$ 
 and that containing $[C_+ A_J A_K A_I]$ 
 represent the same Feynman diagram. 
} %=============
\begin{align}
 \gamma_1: \quad
 &[C_{+} A_I A_J A_K] \,X_{I_1 \ldots I_p, J_1 \ldots J_q, K_1 \ldots K_r}
\nn\\
 &\times
 [Y_{p+1} B_I^{(+)} B_{I_1}^{\sigma_{I_1}} \cdots 
 B_{I_p}^{\sigma_{I_p}} ]  \,
 [Y_{q+1} B_J^{(+)} B_{J_1}^{\sigma_{J_1}} \cdots 
 B_{J_q}^{\sigma_{J_q}} ] \,
 [ Y_{r+1} B_K^{(+)} B_{K_1}^{\sigma_{K_1}} \cdots 
 B_{K_r}^{\sigma_{K_r}} ] 
 \,,  
\label{gamma_1}\\
 \gamma_2: \quad
 &[ C_{+} A_I A_J A_K ]\,X_{I_1 \ldots I_p, J_1 \ldots J_q, K_1 \ldots K_r}
\nn\\
 &\times  [ Y_{p+q+r+3} B_I^{(+)} B_{I_1}^{\sigma_{I_1}} \cdots 
 B_{I_p}^{\sigma_{I_p}} 
 B_K^{(+)} B_{K_1}^{\sigma_{K_1}} \cdots B_{K_r}^{\sigma_{K_r}} 
 B_J^{(+)} B_{J_1}^{\sigma_{J_1}} \cdots B_{J_q}^{\sigma_{J_q}} ]
 \,,
\label{gamma_2} 
\\
 \gamma_3:\quad
 &[ C_{+} A_I A_J A_K ] \,X_{I_1 \ldots I_p, J_1 \ldots J_q, K_1 \ldots K_r}
\nn\\
 &\times  [Y_{p+q+r+3} B_I^{(+)} B_{I_1}^{\sigma_{I_1}} \cdots 
 B_{I_p}^{\sigma_{I_p}} 
 B_J^{(+)} B_{J_1}^{\sigma_{J_1}} \cdots B_{J_q}^{\sigma_{J_q}} 
 B_K^{(+)} B_{K_1}^{\sigma_{K_1}} \cdots B_{K_r}^{\sigma_{K_r}}  ] 
 \,,
\label{gamma_3} 
\\
 \tilde\gamma_1: \quad
 &[ C_{-} A_I A_J A_K ]\,X_{I_1 \ldots I_p, J_1 \ldots J_q, K_1 \ldots K_r}
\nn\\
 &\times 
 [Y_{p+r+2} B_I^{(+)} B_{I_1}^{\sigma_{I_1}} \cdots B_{I_p}^{\sigma_{I_p}} 
 B_K^{(+)} B_{K_1}^{\sigma_{K_1}} \cdots B_{K_r}^{\sigma_{K_r}} ]\,
 [ Y_{q+1} B_J^{(+)} B_{J_1}^{\sigma_{J_1}} \cdots 
 B_{J_q}^{\sigma_{J_q}} ] 
 \,,
\label{tgamma_1} 
\\
\tilde\gamma_2:\quad
 &[ C_{-} A_I A_J A_K ]\,X_{I_1 \ldots I_p, J_1 \ldots J_q, K_1 \ldots K_r}
\nn\\
 &\times 
 [ Y_{q+r+2} B_J^{(+)} B_{J_1}^{\sigma_{J_1}} \cdots 
 B_{J_q}^{\sigma_{J_q}} 
 B_K^{(+)} B_{K_1}^{\sigma_{K_1}} \cdots B_{K_r}^{\sigma_{K_r}} ] \,
 [ Y_{p+1} B_I^{(+)} B_{I_1}^{\sigma_{I_1}} \cdots B_{I_p}^{\sigma_{I_p}} ] 
  \,,
\label{tgamma_2} 
\\
\tilde\gamma_3: \quad
 &[ C_{-} A_I A_J A_K ] \,X_{I_1 \ldots I_p, J_1 \ldots J_q, K_1 \ldots K_r}
\nn\\
 &\times 
 [ Y_{p+q+2} B_I^{(+)} B_{I_1}^{\sigma_{I_1}} \cdots B_{I_p}^{\sigma_{I_p}} 
 B_J^{(+)} B_{J_1}^{\sigma_{J_1}} \cdots B_{J_q}^{\sigma_{J_q}} ]\,
 [ Y_{r+1} B_K^{(+)} B_{K_1}^{\sigma_{K_1}} \cdots 
 B_{K_r}^{\sigma_{K_r}} ]  
  \,. 
\label{tgamma_3}
\end{align}
Here, we have used the abbreviation for Wick contractions 
introduced in subsection \ref{subsec_review_TH}, 
and the superscript $\sigma$ takes $(+)$ or $(-)$. 
We have written explicitly 
only for the part 
of the interaction vertices 
corresponding to $\Delta$ (expressed by $[C_{\pm} A_I A_J A_K]$) 
and the hinges connected to $\Delta$. 
The remaining part 
(denoted by $X_{I_1 \ldots I_p, J_1 \ldots J_q, K_1 \ldots K_r}$) 
is common among the sextet ($\gamma_1, \ldots,  \tilde\gamma_3$) 
and represents the other interaction vertices and their contractions.%
\footnote{%=====
 $X_{I_1 \ldots K_r}$ includes $A_{I_1}$, $\ldots$, 
 $A_{K_r}$, which are the partners of $B_{I_1}$, $\ldots$, $B_{K_r}$ 
 in the Wick contractions.  
} %=============
As for diagrams $\gamma_1,\gamma_2,\gamma_3$, 
the identification at $\Delta$ preserves the orientation 
as in Fig.~\ref{Fig_tetra_id} (a), 
and thus we have used the vertex $[C_{+} A_I A_J A_K]$. 
On the other hand, 
as for diagrams $\tilde\gamma_1,\tilde\gamma_2,\tilde\gamma_3$, 
the identification at $\Delta$ does not preserve the orientation 
as in Fig.~\ref{Fig_tetra_id} (b), 
and thus we have used the vertex $[C_{-} A_I A_J A_K]$. 
In Appendix \ref{sec_sextet_as_tet} 
we prove that the diagrams $\gamma_2$, $\ldots$, $\tilde\gamma_3$ 
represent tetrahedral decompositions 
if $\gamma_1$ does.

So far we have assumed that 
the orientation is preserved at $\Delta$ in $\Gamma_1$ 
and also that $\Gamma_1$ is of the type (1) in \eqref{edge-types}. 
For other cases, 
one can also obtain the corresponding sextets 
$(\gamma_1,\ldots,\tilde\gamma_3)$ 
in a similar way.

%%%%%%%%%%%%%%%%%%%%%%%%%%%%%%%%%%%%%%%%%%%%%
\subsection{Example}
\label{subsec_example}
%%%%%%%%%%%%%%%%%%%%%%%%%%%%%%%%%%%%%%%%%%%%%

\begin{figure}[htbp]
\begin{quote}
\begin{center}
 \includegraphics[height = 3.0cm]{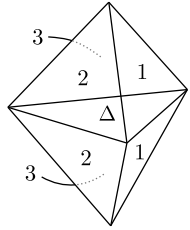}
 \caption
 {
 Tetrahedral decomposition $\Gamma_1$ of three-sphere, 
 where triangles with the same label are identified. 
}
\label{Fig_ex_Gamma_1_tetra}
\vspace{-3ex}
\end{center}
\end{quote}
\end{figure}

\begin{figure}[htbp]
\begin{quote}
\begin{center}
 \includegraphics[height = 5cm]{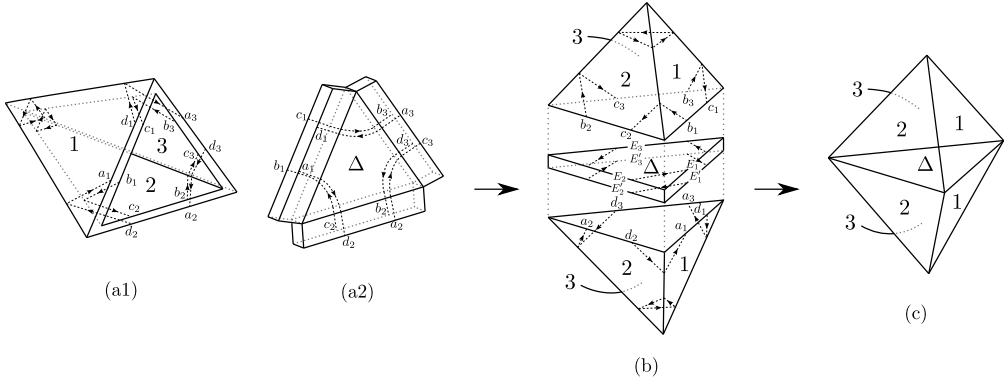}
 \caption{
 Feynman diagram $\gamma_1$ representing $\Gamma_1$. 
 (a1): the part corresponding to $X$. 
 (a2): the part corresponding to $\Delta$. 
%  (i.e.\ the interaction vertices 
%  which appear explicitly in \eqref{ex_gamma_1}). 
 The interaction vertices corresponding to triangles 
 are specified by the labels, $\Delta,1,2,3$. 
 Three 2-hinges exist in (a1) but are not displayed explicitly there. 
 (b): edge-identification at $\Delta$, 
 which is realized by contracting (a1) and (a2). 
 (c): the obtained tetrahedral decomposition $\Gamma_1$. 
 }
\label{Fig_ex_gamma_1}
\vspace{-3ex}
\end{center}
\end{quote}
\end{figure}

To understand the meaning of the above sextet 
\eqref{gamma_1}--\eqref{tgamma_3}, 
let us consider a simple example. 
We take a tetrahedral decomposition $\Gamma_1$ 
of a three-sphere, 
consisting of two tetrahedra glued together at their faces 
as shown in Fig.~\ref{Fig_ex_Gamma_1_tetra}. 
%%%%%%%%%%%%%%%%%%%%%%%
%%%%%%%%%%%%%%%%%%%%%%%
Diagram $\gamma_1$ representing $\Gamma_1$ is realized 
by the following group of Wick contractions [eq.~\eqref{gamma_1}]: 
\begin{align}
 &[C_{+}A_I A_J A_K ]_\Delta [Y_2 B_I^{(+)} B_{I_1}^{(-)} ] 
 [Y_2 B_J^{(+)} B_{J_1}^{(-)} ] [Y_2 B_K^{(+)} B_{K_1}^{(-)} ] 
 \times X_{I_1 J_1 K_1} \,, 
\label{gamma_1_example}
\\ 
 & X_{I_1 J_1 K_1} = [C_{+} A_{I_1} A_{M_1} A_{N_2}]_1 [C_{+} A_{J_1} 
 A_{N_1} A_{L_2} ] _2 [C_{+} A_{K_1} A_{L_1} A_{M_2} ]_3  
\nn\\
 &\qquad\quad \quad \quad \times  [Y_2 B_{L_1}^{(+)} B_{L_2}^{(-)} ] 
 [Y_2 B_{M_1}^{(+)} B_{M_2}^{(-)} ] [Y_2 B_{N_1}^{(+)} B_{N_2}^{(-)} ] \,.  
\label{ex_gamma_1}
\end{align}
Here, the subscripts $\Delta,1,2,3$ specify the triangles 
corresponding to the interaction vertices; 
$\Delta$ specifies the triangle 
at which we change the edge-identification 
to obtain $\gamma_2, \ldots, \tilde\gamma_3$, 
and $1,2,3$ specify the triangles 
belonging to the rest part $X$, 
which consists of three triangles ($1,2,3$) and three 2-hinges 
(see Fig.~\ref{Fig_ex_gamma_1}). 
%%%%%%%%%%%%%%%%%%%%%%%
%%%%%%%%%%%%%%%%%%%%%%%
Fig.~\ref{Fig_ex_gamma_1} (a1) depicts the part corresponding to $X$, 
while Fig.~\ref{Fig_ex_gamma_1} (a2) depicts 
the part corresponding to $\Delta$, 
which consists of a triangle and three 2-hinges. 
In Fig.~\ref{Fig_ex_gamma_1} (a1) and (a2), 
edges with the same indices are connected by contractions. 
Recalling that the edge-identification of $\Gamma_1$ 
is expressed by $(E_1 E_2 E_3)=(E'_1 E'_2 E'_3)$, 
we label the ordered indices 
$(b_i, c_i)$ and $(a_i,d_i)$ $(i=1,2,3)$ in Fig.~\ref{Fig_ex_gamma_1} (a1) 
as $E_i=E_i(b_i,c_i)$ and $E'_i=E'_i(a_i,d_i)$, respectively. 
Since edges $E_i$ and $E'_i$ are expressed as in Fig.~\ref{Fig_ex_gamma_1} (b) 
when (a1) and (a2) are combined, 
we see that the edge-identification $(E_1 E_2 E_3)=(E'_1 E'_2 E'_3)$ 
will certainly be realized 
after triangle $\Delta$ is deflated.

\begin{figure}[htbp]
\begin{quote}
\begin{center}
 \includegraphics[height = 5cm]{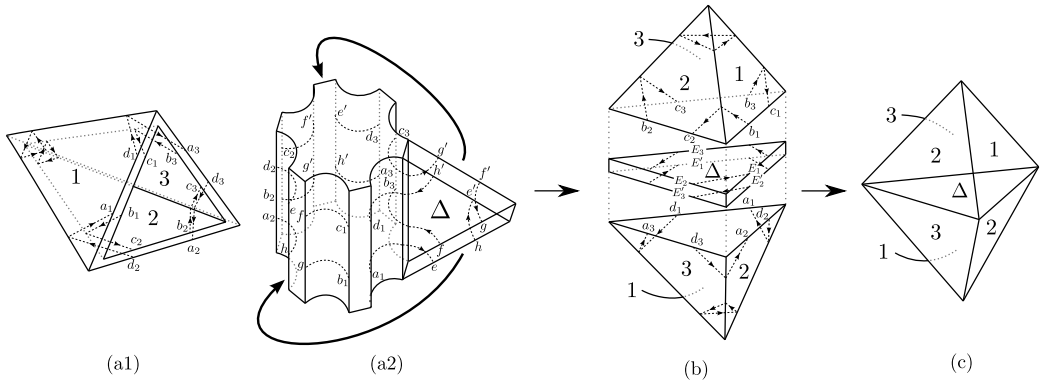}
 \caption{
 Feynman diagram $\gamma_2$ representing $\Gamma_2$. 
 (a1): the part corresponding to $X$. 
 (a2): the part corresponding to $\Delta$. 
 (b): the edge-identification at $\Delta$. 
 (c): the obtained tetrahedral decomposition $\Gamma_2$. 
 }
\label{Fig_ex_gamma_2}
\vspace{-3ex}
\end{center}
\end{quote}
\end{figure}

Now we consider diagram $\gamma_2$ 
representing the tetrahedral decomposition $\Gamma_2$ 
that is obtained from $\Gamma_1$ 
by changing the edge-identification 
from $(E_1 E_2 E_3)=(E'_1 E'_2 E'_3)$
to $(E_1 E_2 E_3)=(E'_2 E'_3 E'_1)$ 
(see Fig.~\ref{Fig_ex_gamma_2}). 
%%%%%%%%%%%%%%%%%%%%%%%
%%%%%%%%%%%%%%%%%%%%%%%
Note that $\Gamma_2$ has 
the topology of a three-dimensional lens space $L(3,1)$, 
as can be seen from Fig.~\ref{Fig_ex_gamma_2} (c). 
Diagram $\gamma_2$ is given by \eqref{gamma_2}, that is, 
\begin{align}
 &[C_{+}A_I A_J A_K ] [Y_6 B_I^{(+)}  B_{I_1}^{(-)} B_K^{(+)} 
 B_{K_1}^{(-)} B_J^{(+)} B_{J_1}^{(-)} ] \times X_{I_1 J_1 K_1} \,. 
\label{ex_gamma_2} 
\end{align}
Since the part given by $X$ is common among the sextet, 
we have the same labeling of edges, 
$E_i(b_i,c_i)$ and $E'_i(a_i,d_i)$ $(i=1,2,3)$, 
and diagram $\gamma_2$ takes the form 
shown in Fig.~\ref{Fig_ex_gamma_2} (a1) and (a2). 
In Fig.~\ref{Fig_ex_gamma_2} (a2), 
the ordered indices $(b_1, c_1)$ 
are connected to $(g,f)$ by index lines, 
and $(a_2, d_2)$ are connected to $(h,e)$. 
Thus, as can be seen from Fig.~\ref{Fig_ex_gamma_2} (b), 
edge $E_1(b_1,c_1)=(g, f)$ will be identified 
with edge $E'_2(a_2,d_2)=(h,e)$ 
after triangle $\Delta$ is deflated.   
% Thus, the edge with indices $(b_1, c_1)$ in Fig.~\ref{Fig_ex_gamma_2} (a1) 
% is identified with that with $(a_2, d_2)$ in Fig.~\ref{Fig_ex_gamma_2} (a1) 
% through the part given by $\Delta$, 
% that is, the edge $E_1$ is identified with the edge $E_2^\prime$. 
Similarly, edges $E_2(b_2, c_2)$ and $E_3(b_3, c_3)$ 
will be identified with edges $E'_3(a_3, d_3)$ and $E'_1(a_1, d_1)$, 
respectively. 
We thus obtain the edge-identification 
$(E_1 E_2 E_3)=(E'_2 E'_3 E'_1)$ of $\Gamma_2$. 
% This means the edge-identification 
% $(E_1 E_2 E_3) = (E_2^\prime E_3^\prime E_1^\prime)$. 
% The diagram certainly represents the tetrahedral decomposition 
% $\Gamma_2$ (see Fig.~\ref{Fig_ex_gamma_2} (b) and (c)). 
In a similar way, 
we can realize $\Gamma_3$ 
[resulting from the edge-identification 
$(E_1 E_2 E_3) = (E_3^\prime E_1^\prime  E_2^\prime)$ at $\Delta$]
as a diagram $\gamma_3$ [eq.~\eqref{gamma_3}] of a triangle-hinge model. 
$\Gamma_3$ has the topology of a lens space $L(3,2) = L(3,1)$. 
Thus, in this simple example, 
two diagrams $\gamma_2$ and $\gamma_3$ represent 
the same tetrahedral decomposition, $\Gamma_2 = \Gamma_3$.

\begin{figure}[htbp]
\begin{quote}
\begin{center}
 \includegraphics[height = 5cm]{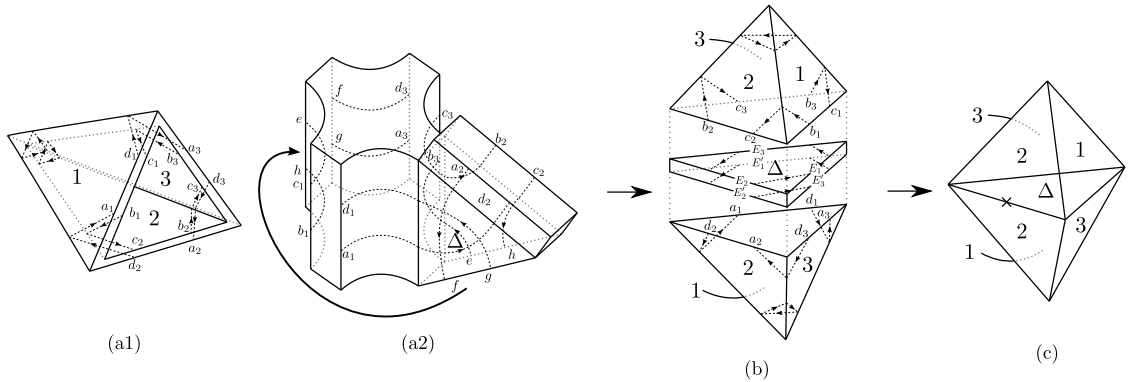}
 \caption{
 Feynman diagram $\tilde\gamma_1$ representing $\tilde\Gamma_1$. 
 (a1): the part corresponding to $X$. 
 (a2): the part corresponding to $\Delta$. 
 (b): the edge-identification at $\Delta$. 
 (c): the obtained tetrahedral decomposition $\tilde\Gamma_1$,  
 where the midpoint of the edge shared 
 by triangles 2 and $\Delta$ is singular. 
 }
\label{Fig_ex_gamma_t1}
\vspace{-3ex}
\end{center}
\end{quote}
\end{figure}

As for the tetrahedral decomposition $\tilde\Gamma_1$  
[resulting from the edge-identification 
$(E_1 E_2 E_3) = (E'_3 E'_2 E'_1)$ at $\Delta$], 
the corresponding diagram $\tilde\gamma_1$ 
is obtained from the following group of Wick contractions 
[eq.~\eqref{tgamma_1}]:
\begin{align}
 &[C_{-} A_I A_J A_K] [ Y_4 B_I^{(+)} B_{I_1}^{(-)} B_K^{(+)} 
 B_{K_1}^{(-)} ] [Y_2 B_J^{(+)} B_{J_1}^{(-)} ] 
 \times X_{I_1 J_1 K_1} \,.  
\label{ex_gamma_t1}
\end{align}
The diagram is depicted in Fig.~\ref{Fig_ex_gamma_t1}. 
%%%%%%%%%%%%%%%%%%%%%%%
%%%%%%%%%%%%%%%%%%%%%%%
In Fig.~\ref{Fig_ex_gamma_t1} (a2), 
the ordered indices $(b_1, c_1)$ 
are connected to $(h, e)$ by index lines, 
and $(d_3, a_3)$ are connected to $(f, g)$. 
Thus, due to the edge-identification for $C_-$ 
explained below \eqref{cminus}, 
edge $E_1(b_1,c_1)=(h, e)$ 
will be identified with edge $E'_3(d_3,a_3)=(f, g)$ 
after triangle $\Delta$ is deflated. 
Similarly, edges $E_2(b_2, c_2)$ and $E_3(b_3, c_3)$ 
will be identified with edges $E'_2(d_2, a_2)$ and $E'_1(d_1, a_1)$, 
respectively. 
We thus obtain the edge-identification 
$(E_1 E_2 E_3) = (E'_3 E'_2 E'_1)$ of $\tilde\Gamma_1$. 
% due to the edge-identification of $C_-$ explained 
% in subsection \ref{subsec_action_tet}, 
% the edge with indices $(b_1, c_1)$ in Fig.~\ref{Fig_ex_gamma_t1} (a1) 
% will be identified with that with $(d_3, a_3)$ 
% in Fig.~\ref{Fig_ex_gamma_t1} (a1) through the part given by $\Delta$, 
% that is, edge $E_1$ will be identified with edge $E_3^\prime$. 
% Similarly, edges $E_2(b_2, c_2)$ and $E_3(b_3, c_3)$ 
% will be identified with edges $E'_1(d_1, a_1)$ and $E'_2(d_2, a_2)$, 
% respectively. 
% This means the edge-identification 
% $(E_1 E_2 E_3) = (E_3^\prime E_2^\prime E_1^\prime)$. 
% The diagram certainly represents the tetrahedral decomposition 
% $\tilde\Gamma_1$ (see Fig.~\ref{Fig_ex_gamma_t1} (b) and (c)). 
Although $\tilde\Gamma_1$ consists of two tetrahedra, 
it is not a three-dimensional manifold. 
In fact, there is a singularity at the midpoint of edge 
$E_2 = E_2^\prime$ of $\Delta$, 
around which we cannot define a local orientation. 
It is easy to see that the other diagrams $\tilde\gamma_2$ 
and $\tilde\gamma_3$ are realized 
by \eqref{tgamma_2} and \eqref{tgamma_3}, respectively, 
and represent the same tetrahedral decomposition 
as $\tilde\Gamma_1$.

%%%%%%%%%%%%%%%%%%%%%%%%%%%%%%%%%%%%%%%%%%%%%
\subsection{Note on the weights of diagrams}
\label{sec_same_weight}
%%%%%%%%%%%%%%%%%%%%%%%%%%%%%%%%%%%%%%%%%%%%%

We comment that 
a sextet $(\gamma_1,\gamma_2,\gamma_3,\tilde\gamma_1, 
\tilde\gamma_2, \tilde\gamma_3)$ 
appears in the free energy with the same coefficients. 
By ``the same coefficients'' 
we mean that the numerical factors of these diagrams are the same 
except for powers of $n, \lambda, \mu_k$
if we treat the common part $X$ as a set of distinguished external vertices 
and sum over all diagrams representing the same tetrahedral decomposition.

We note that, 
if a group $x$ of Wick contractions represents 
a tetrahedral decomposition  
and if all the interaction vertices are distinguished, 
then $x$ contributes to the free energy as%
\footnote{%=====
 The $n$ dependence comes from the assumption 
 that all the index polygons are triangles.
} %=============
\begin{align}
 \frac{1}{s_2!}  \Bigl(\frac{\lambda}{6n} \Bigr)^{s_2}  
 \prod_{k=1}\biggl[ \frac{1}{s_1^k!}  
 \Bigl(\frac{n^2 \mu_k}{2k} \Bigr)^{s_1^k} 
 \biggr] \,. 
\label{distinguished_contraction}
\end{align}
Here, $s_2$ and $s_1^k$ denote the numbers of triangles and $k$-hinges, 
respectively, in diagram $\gamma=[x]$. 
Thus, there arise $1/(s_2! \, 6^{s_2})$ and 
$1/(s_1^k! \, (2k)^{s_1^k})$ in the free energy as numerical factors.

If there are $n_k$ internal $k$-hinges in a diagram,%
\footnote{%=====
 Internal hinges mean the parts not in $X$  
 but connected to $\Delta$.
} %=============
there are $n_k!$ different contractions corresponding to 
the permutation of these hinges, 
since the external vertices are distinguished. 
For each $k$-hinge, 
there are $2k$ ways to give the same diagram 
due to the symmetry of $k$-hinge vertices.  
Thus, the numerical factor of each contraction, $1/(n_k! (2k)^{n_k})$, 
is compensated 
if we sum these contributions. 
The numerical factor $1/6$ of triangle $\Delta$ is also canceled.  
Actually, since $C_+$ has the symmetry \eqref{symmetry_C} 
there are six ways to give the same diagram.  
The above computation ensures that  
three diagrams 
$\gamma_1,\gamma_2,\gamma_3$ 
are generated with unit coefficient in the original triangle-hinge model. 
Furthermore,  
since $C_-$ also has the symmetry 
\begin{align}
 C_-^{i_1 j_1 i_2 j_2 i_3 j_3} = C_-^{i_2 j_2 i_3 j_3 i_1 j_1}
 = C_-^{j_1 i_1 j_2 i_2 j_3 i_3}\,,
\label{symmetry_C_-}
\end{align} 
$\tilde\gamma_1, \tilde\gamma_2, \tilde\gamma_3$ 
are also generated with unit coefficient in an unoriented model.

%%%%%%%%%%%%%%%%%%%%%%%%%%%%%%%%%%%%%%%%%%%%%
%%%%%%%%%%%%%%%%%%%%%%%%%%%%%%%%%%%%%%%%%%%%%
\section{Matter fields in unoriented triangle-hinge models}
\label{sec_matter}
%%%%%%%%%%%%%%%%%%%%%%%%%%%%%%%%%%%%%%%%%%%%%
%%%%%%%%%%%%%%%%%%%%%%%%%%%%%%%%%%%%%%%%%%%%%

In this section we show that matter fields  
can be introduced to unoriented triangle-hinge models 
in the same way as the original triangle-hinge models 
\cite{Fukuma:2015haa}. 
We here focus on assigning matter degrees of freedom only to tetrahedra, 
but the assignment can be done to simplices of any dimensions 
as in \cite{Fukuma:2015haa}. 

\begin{figure}[htbp]
\begin{quote}
\begin{center}
 \includegraphics[height = 3cm]{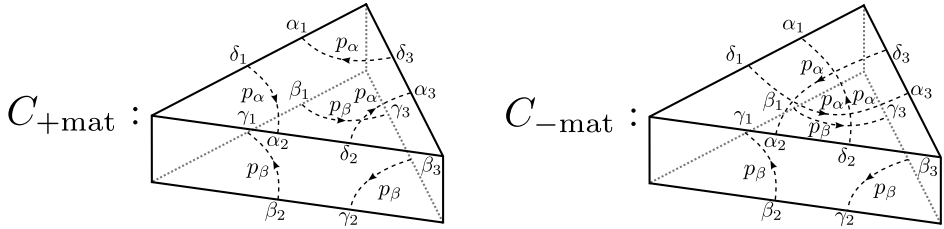}
 \caption{Interaction vertices corresponding to triangle 
 ($\mathcal{A}_{\rm mat}$ part). 
 The upper (lower) side of each triangle has color $\alpha$ ($\beta$).}
\label{Fig_3simplex_coloring}
\vspace{-3ex}
\end{center}
\end{quote}
\end{figure}

Introducing matter degrees of freedom is realized 
by coloring each tetrahedron in tetrahedral decompositions. 
Actually, we only need to repeat the steps given in \cite{Fukuma:2015haa}. 
We first extend the algebra $\mathcal{A}$ to a tensor product of the form 
$\mathcal{A} = \mathcal{A}_{\rm grav} \otimes \mathcal{A}_{\rm mat}$. 
Here, $\mathcal{A}_{\rm grav}$ is again $M_{n=3m}(\mathbb{R})$, 
and we take $\mathcal{A}_{\rm mat}$ to be $M_{|\mathcal{J}|}(\mathbb{R})$, 
where $\mathcal{J}$ is the set of colors. 
Now the dynamical variables $A$, $B$ have eight indices 
$A= (A_{a b \alpha \beta, c d \gamma \delta})$, 
$B= (B_{a b \alpha \beta, c d \gamma \delta})$, 
where the indices $a,b,c,d$ correspond to $\mathcal{A}_{\rm grav}$, 
and $\alpha,\beta,\gamma,\delta$ to $\mathcal{A}_{\rm mat}$.%
\footnote{%=====
 $A$ and $B$ are real-valued matrices 
 symmetric with respect to the pair of indices, 
 $A_{a b \alpha \beta, c d \gamma \delta} 
 = A_{c d \gamma \delta, a b \alpha \beta}$, 
 $B_{a b \alpha \beta, c d \gamma \delta} 
 = B_{c d \gamma \delta, a b \alpha \beta}$. 
} %=============
Next we set the tensor $C$ to take the form $C = C_{+} + C_{-}$ 
($C_\pm$ represents two ways to glue tetrahedra 
depicted in Fig.~\ref{Fig_tri_+-}), 
and assume that each has a factorized form 
$C_{\pm} = C_{\pm \rm grav} C_{\pm \rm mat}$. 
Here, we set $C_{\pm \rm grav}$ to the form \eqref{unoriented_THaction}, 
and let $C_{\pm \rm mat}$ take the following form: 
\begin{align}
 & C_{+ \rm mat}^{\alpha_1 \beta_1 \gamma_1 \delta_1
 \alpha_2 \beta_2 \gamma_2 \delta_2 \alpha_3 \beta_3 \gamma_3 \delta_3} 
 = \sum_{\alpha, \beta \in \mathcal{J}} \lambda_{\alpha \beta} \,
 p_{\alpha}^{\delta_1 \alpha_2}p_{\alpha}^{\delta_2 \alpha_3}
 p_{\alpha}^{\delta_3 \alpha_1} 
 p_{\beta}^{\beta_3 \gamma_2}p_{\beta}^{\beta_2 \gamma_1}
 p_{\beta}^{\beta_1 \gamma_3}\,,
\label{colored_C+}
\\
& C_{- \rm mat}^{\alpha_1 \beta_1 \gamma_1 \delta_1 
 \alpha_2 \beta_2 \gamma_2 \delta_2 \alpha_3 \beta_3 \gamma_3 \delta_3} 
 = \sum_{\alpha, \beta \in \mathcal{J}} \lambda_{\alpha \beta} \,
 p_{\alpha}^{\delta_3 \alpha_2}p_{\alpha}^{\delta_2 \alpha_1}
 p_{\alpha}^{\delta_1 \alpha_3} 
 p_{\beta}^{\beta_3 \gamma_2}p_{\beta}^{\beta_2 \gamma_1}
 p_{\beta}^{\beta_1 \gamma_3}\,, 
\label{colored_C-}
\end{align}
where 
$p_\alpha=(p_\alpha^{\beta \gamma} 
= \delta_\alpha^\beta \delta_\alpha^\gamma)$ 
is the projector to the $\alpha$-th component  
and $\lambda_{\alpha \beta}$ is a real constant. 
Equations \eqref{colored_C+} and \eqref{colored_C-} 
mean that we insert $p_\alpha$ to each index lines 
(as $\omega$ was inserted for $\mathcal{A}_{\rm grav}$) 
and take a summation over $\alpha$ and $\beta$ 
with weight $\lambda_{\alpha \beta}$. 
The $\mathcal{A}_{\rm mat}$ part of the interaction vertex 
$[C_{+}A^3] + [C_{-}A^3]$ 
can be illustrated as in Fig.\ref{Fig_3simplex_coloring}. 
%%%%%%%%%%%%%%%%%%%%%%%%%%%%%%%%%%%%%%%%%%%
%%%%%%%%%%%%%%%%%%%%%%%%%%%%%%%%%%%%%%%%%%%
Note that $p_\alpha$ is common among three index lines 
on each side of a triangle. 
Thus, one can say that each side of triangle has a color.

\begin{figure}[htbp]
\begin{quote}
\begin{center}
 \includegraphics[height = 4cm]{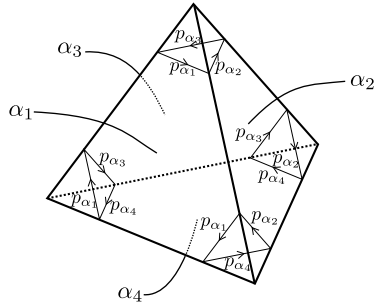}
 \caption{Index triangles inside a tetrahedron with triangles colored 
 as in \eqref{colored_C+},\eqref{colored_C-}\cite{Fukuma:2015haa}.
 }
\label{Fig_colored_tetra}
\vspace{-3ex}
\end{center}
\end{quote}
\end{figure}

In this construction, the index function $\mathcal{F}(\gamma)$ 
of diagram $\gamma$ is factorized to the form
\begin{align}
 \mathcal{F}(\gamma) \equiv \mathcal{F}(\gamma; \mathcal{A}) 
 = \mathcal{F}(\gamma; \mathcal{A}_{\rm grav}) 
 \mathcal{F}(\gamma; \mathcal{A}_{\rm mat}) 
 \equiv \mathcal{F}_{\rm grav}(\gamma) \mathcal{F}_{\rm mat}(\gamma), 
\end{align}
and the factor $\mathcal{F}_{\rm grav}(\gamma)$ ensures 
that diagram $\gamma$ represents a tetrahedral decomposition. 
The index lines corresponding to $\mathcal{A}_{\rm mat}$ 
also form index triangles (see Fig.\ref{Fig_colored_tetra}). 
%%%%%%%%%%%%%%%%%%%%%%%%%%%%%%%%%%%%%%%%%%%
%%%%%%%%%%%%%%%%%%%%%%%%%%%%%%%%%%%%%%%%%%%
A tetrahedron surrounded by four side of triangles with color 
$\alpha_1,\alpha_2,\alpha_3,\alpha_4$ gives the factor 
\begin{align}
 &\mathrm{tr}(p_{\alpha_1} p_{\alpha_2} p_{\alpha_3}) 
 \mathrm{tr}(p_{\alpha_2} p_{\alpha_1} p_{\alpha_4}) 
 \mathrm{tr}(p_{\alpha_1} p_{\alpha_3} p_{\alpha_4}) 
 \mathrm{tr}(p_{\alpha_3} p_{\alpha_2} p_{\alpha_4}) \nonumber \\
 &= \left\{\begin{array}{l}
 1 \quad (\alpha_1 = \alpha_2 = \alpha_3 = \alpha_4) \\
 0 \quad ({\rm otherwise}). 
 \end{array}\right.
\end{align}
This means that $\mathcal{F}_{\rm mat}(\gamma)$ can take 
nonvanishing values 
only when four side of triangles of each tetrahedron 
have the same color (say $\alpha$), 
which enables us to say that the tetrahedron has the color $\alpha$. 
We thus succeed in coloring tetrahedra in $\gamma$.

If two tetrahedra with color $\alpha$ and $\beta$ are glued at their faces, 
the face (corresponding to $C_{\pm \rm mat}$) gives the factor  
$\lambda_{\alpha \beta}$. 
In this sense the coupling constants $\lambda_{\alpha \beta}$ define 
a local interaction between the color $\alpha$ and $\beta$ 
\cite{Fukuma:2015haa}. 
If we take the set of colors, $\mathcal{J}$, to be 
$\mathbb{R}^D = \{\bf x\}$ 
and let the coupling constants 
$\lambda_{\bf x,y}$ (${\bf x,y} \in \mathbb{R}^D$) take nonvanishing values 
only around $\bf y$ as a function of $\bf x$, 
then $\bf x$ can be interpreted as the target space coordinates 
of a tetrahedron in $\mathbb{R}^D$. 
Since neighboring tetrahedra are locally connected in $\mathbb{R}^D$, 
the model can describe the dynamics of unoriented membranes embedded 
in $\mathbb{R}^D$.

%%%%%%%%%%%%%%%%%%%%%%%%%%%%%%%%%%%%%%%%%%%%%
%%%%%%%%%%%%%%%%%%%%%%%%%%%%%%%%%%%%%%%%%%%%%
\section{Conclusion and outlook}
\label{sec_conclusion}
%%%%%%%%%%%%%%%%%%%%%%%%%%%%%%%%%%%%%%%%%%%%%
%%%%%%%%%%%%%%%%%%%%%%%%%%%%%%%%%%%%%%%%%%%%%

In this paper, 
we first defined unoriented membrane theories 
in terms of tetrahedral decompositions, 
and then realized them as triangle-hinge models. 
Unoriented membrane theories are obtained 
from oriented open membrane theories of disk topology  
by gauging the worldvolume parity transformation $\Omega$. 
For each triangle $\Delta$ in a tetrahedral decomposition,  
we have introduced two types of triplets, $(\Gamma_1,\Gamma_2,\Gamma_3)$ 
and $(\tilde\Gamma_1, \tilde\Gamma_2, \tilde\Gamma_3)$,  
which respectively correspond to two ways of identification at $\Delta$, 
\eqref{++} and \eqref{+-}. 
The transformation $\Omega$ is then defined as the interchange 
between $(\Gamma_1,\Gamma_2,\Gamma_3)$ 
and $(\tilde\Gamma_1, \tilde\Gamma_2, \tilde\Gamma_3)$. 
After gauging $\Omega$, 
an unoriented membrane theory treats 
all the tetrahedral decompositions in the sextet 
$(\Gamma_1,\Gamma_2,\Gamma_3, 
\tilde\Gamma_1, \tilde\Gamma_2, \tilde\Gamma_3)$ 
equally.

An unoriented membrane theory is realized 
as a triangle-hinge model with the action \eqref{unoriented_THaction}. 
It generates Feynman diagrams 
representing unoriented tetrahedral decompositions. 
We gave explicitly in \eqref{gamma_1}--\eqref{tgamma_3} 
the sextet of Feynman diagrams 
$(\gamma_1, \ldots, \tilde\gamma_3)$ 
corresponding to $(\Gamma_1, \ldots, \tilde\Gamma_3)$, 
and showed that these six diagrams appear with unit coefficient 
up to factors of coupling constants 
if we treat the common part $X$ 
as a set of distinguished external vertices 
and sum over all Wick contractions giving the same diagram. 
We further showed that 
matter degrees of freedom can be introduced 
to unoriented triangle-hinge models 
by coloring tetrahedra 
as carried out in \cite{Fukuma:2015haa}.  
Although we only discussed the coloring of tetrahedra in this paper, 
we can set matter degrees of freedom to simplices of any dimensions 
(i.e.\ tetrahedra, triangles, edges and/or vertices)
as in \cite{Fukuma:2015haa}.

We expect that unoriented triangle-hinge models are solvable 
at least at the same level of the original oriented models \cite{FSU_prep}, 
since the dynamical variables are the same type of matrices 
and the actions have almost the same structure 
as the original oriented triangle-hinge models. 
The unoriented models actually might be easier to solve 
than the original oriented models, 
because the interaction term corresponding to a triangle 
has higher symmetry, 
which may help us to carry out the path-integrals more analytically. 
It is interesting to study critical behaviors of the models 
in both analytical and numerical ways.

%%%%%%%%%%%%%%%%%%%%%%%%%%%%%%%%%%%%%%%
%%%%%%%%%%%%%%%%%%%%%%%%%%%%%%%%%%%%%%%
\section*{Acknowledgments}
The authors thank Naoki Sasakura 
for useful discussions. 
MF is supported by MEXT (Grant No.\,23540304).
SS is supported by the JSPS fellowship.
%%%%%%%%%%%%%%%%%%%%%%%%%%%%%%%%%%%%%%%
%%%%%%%%%%%%%%%%%%%%%%%%%%%%%%%%%%%%%%%

\appendix

%%%%%%%%%%%%%%%%%%%%%%%%%%%%%%%%%%%%%%%%%%%%%
\section{Sextet as tetrahedral decompositions}
\label{sec_sextet_as_tet}
%%%%%%%%%%%%%%%%%%%%%%%%%%%%%%%%%%%%%%%%%%%%%

In this Appendix, 
we show that 
the diagrams $\gamma_2$, $\ldots$, $\tilde\gamma_3$ 
represent tetrahedral decompositions 
if $\gamma_1$ does. 
Let us look into the indices 
in the group of Wick contractions \eqref{gamma_1}. 
We label the indices of $B_{I_1}$ 
as $B_{a'_{I_1} b'_{I_1} c'_{I_1} d'_{I_1}}$,  
while we label those of $A_{I_1}$ as 
$A_{a_{I_1} b_{I_1} c_{I_1} d_{I_1}}$ 
or $A_{ c_{I_1} d_{I_1} a_{I_1} b_{I_1}}$ 
according to $\sigma_{I_1}=(+)$ or $(-)$, 
so that we always have 
\begin{align}
 \contraction{}{A}{_{I_1}}{B} A_{I_1} B_{I_1}^{\sigma_{I_1}} 
 = \delta_{a_{I_1} a'_{I_1}} \delta_{b_{I_1} b'_{I_1}}  
 \delta_{c_{I_1} c'_{I_1}} \delta_{d_{I_1} d'_{I_1}} \,.
\label{contraction_I_1}
\end{align}
We use a similar labeling for other $A_{I_i}$ and $B_{I_i}$.  
Through the parts other than $X$ in \eqref{gamma_1}, 
indices $a_{I_1}$, $\ldots$, $d_{K_r}$ of $X$ are connected 
to each other by index lines. 
The other parts are given by
\begin{align}
 [C_{+} A_I A_J A_K] 
 &= A_{a_I b_I c_I d_I} A_{a_J b_J c_J d_J} A_{a_K b_K c_K d_K} 
 \omega^{d_I a_J} \omega^{d_J a_K} \omega^{d_K a_I} 
 \omega^{b_K c_J} \omega^{b_J c_I} \omega^{b_I c_K} \,,  
\label{c+_contraction}
\\
 [Y_{p+1} B_I B_{I_1} \cdots B_{I_p}] &= B_{a'_I b'_I c'_I d'_I} 
 B_{a'_{I_1}  b'_{I_1} c'_{I_1} d'_{I_1}} \cdots B_{a'_{I_p} b'_{I_p} 
 c'_{I_p} d'_{I_p}} 
 \delta_{b'_I a'_{I_1}}\cdots  \delta_{b'_{I_p} a'_I} 
 \delta_{c'_I d'_{I_1}} \cdots \delta_{c'_{I_p} d'_I} \,. 
\end{align}
Thus, by combining them with \eqref{contraction_I_1}, 
the index lines connecting $a_{I_1}$, $\ldots$, $d_{K_r}$ are given by 
\begin{align}
 \omega^{c_{I_p} b_{J_q}} \omega^{c_{J_q} b_{K_r}} \omega^{c_{K_r} b_{I_p}} 
 \omega^{a_{K_1} d_{J_1}} \omega^{a_{J_1} d_{I_1}} \omega^{a_{I_1} d_{K_1}}
 \delta_{b_{I_1} a_{I_2}} \cdots \delta_{b_{I_{p-1}} a_{I_p}} 
 \delta_{c_{I_1} d_{I_2}} \cdots \delta_{c_{I_{p-1}} d_{I_p}} \cdots \,. 
\label{index_line_gamma_1}
\end{align}
One can find that index lines out of $X$ 
take the same form as \eqref{index_line_gamma_1} 
in diagrams $\gamma_2$, $\ldots$, $\tilde\gamma_3$. 
Since we assume that diagram $\gamma_1$ represents 
tetrahedral decomposition $\Gamma_1$, 
all the index loops in $\gamma_1$ make index triangles. 
Then, the index loops of $\gamma_2$, $\ldots$, $\tilde\gamma_3$  
also must make only index triangles 
because the index loops have the same form as those of $\gamma_1$.  
We thus have shown that the diagrams represent tetrahedral decompositions 
if $\gamma_1$ does.%

%%%%%%%%%%%%%%%%%%%%%%%%%%%%%%%%%%%%%%%
%%%%%%%%%%%%%%%%%%%%%%%%%%%%%%%%%%%%%%%

%%%%%%%%%%%%%%%%%%%%%%%%%%%%%%%%%%%%%%%
%%%%%%%%%%%%%%%%%%%%%%%%%%%%%%%%%%%%%%%
\end{document}